\documentclass[11pt]{article}
\usepackage[english]{babel}
\usepackage{geometry}
\usepackage{float}
\usepackage{mathrsfs}
\usepackage{bm}
\usepackage{amsmath}
\usepackage{amssymb}
\usepackage{graphicx}
\usepackage[round]{natbib}
\usepackage{fullpage}
\usepackage{layout}
\usepackage{multirow}
\usepackage{multicol}
\usepackage{pdflscape}
\usepackage{setspace}
\usepackage{multicol}
\usepackage{subfig}

\usepackage{amsthm}
\usepackage{graphicx}

\begin{document}

\title{A Bayesian Semiparametric Vector Multiplicative Error Model}
\author{ Nicola Donelli \thanks{CGnal s.r.l., Research and Development Department}
	\and
	Stefano Peluso\thanks{Corresponding author. Universit\`{a} degli Studi di Milano-Bicocca, Department of Statistics and Quantitative Methods, via Bicocca degli Arcimboldi 8, 20126 Milan (Italy). E-mail: stefano.peluso@unimib.it}
	\and
	Antonietta Mira \thanks{Universit\`{a} della Svizzera italiana, Institute of Computational Science and Universit\`{a} dell'Insubria, Department of Science and High Technology.}}
\maketitle

\onehalfspacing

\begin{abstract}
Interactions among multiple time series of positive random variables are crucial in diverse financial applications, from spillover effects to volatility interdependence.
A popular model in this setting is the vector Multiplicative Error Model (vMEM) which poses a linear iterative structure on the dynamics of the conditional mean, perturbed by a multiplicative innovation term. A main limitation of vMEM is however its restrictive assumption on the distribution of the random innovation term. A Bayesian semiparametric approach that models the innovation vector as an infinite location-scale mixture of multidimensional kernels with support on the positive orthant is used to address this major shortcoming of vMEM. Computational complications arising from the constraints to the positive orthant are avoided through the formulation of a slice sampler on the parameter-extended unconstrained version of the model. The method is applied on simulated and real data and a flexible specification is obtained that outperforms the classical ones in terms of fitting and predictive power.
\end{abstract}

\section{Introduction}

In many fields of application it is of interest to study the interactions between time series of random variables constrained to be positive, for instance with high-frequency financial variances, durations and trading volumes.
A class of models which is particularly suited to non-negative time series are the \textit{Multiplicative Error Models} (MEMs) introduced by \citet{Engle2002NewFrontiersForARCH} to overcome some drawbacks associated with the conventional workarounds used to model non-negative data such as ignoring the non-negativity constraint or taking the logs of the observations. The advantages of modelling a  non-negative process using distributions with support on the non-negative orthant are well described by \citet{engle1998autoregressive} and can be summarized in these two stylized facts: these distributions do not require complex constraints on their moments to reduce the probability of obtaining negative values during sampling processes and they handle naturally exact zeros (i.e. without adding small constants before taking the logs, that could affect considerably the outcomes).
Specifications of MEMs ask for specific forms of the dynamics in the conditional mean and for the distribution of the multiplicative error term. Diverse choices lead to models of different generality: a scale-factor structure of the conditional mean and a Gamma-distributed error term reduce MEM to the non-negative GARCH model of \citet{Engle2002NewFrontiersForARCH}, whilst an autoregressive structure on the conditional expectation and a Weibull error term lead to the ACD model of \citet{engle1998autoregressive}. The univariate setting has been extended by \citet{Engle2006Multiple} to multivariate time series, with the purpose of analysing the interdependence across volatility measures. In the multivariate version of MEM, \textit{vector MEM} or vMEM, the vector of observations at a given time is represented as the element-by-element product of a conditional mean and a random innovation, with some assumed multidimensional distribution for the innovation vector and a specified dynamics for the conditional mean vector. 
\citet{Engle2006Multiple} initially proposed a natural extension of the univariate case, assuming that the components of the innovation vector are independently distributed with Gamma densities with equal shape and rate. The same approach has been followed by \citet{Engle2012VolatilitySpilloversInAsia} and \citet{Giovannetti2011MEM} to analyse the spillover effect between different volatility proxies. To weaken the assumption of stochastic independence among the components of the innovation vector, \citet{Cipollini2006vector} explored the possibility to use multivariate Gamma distributions defined on the positive orthant \citep{Johnson2000ContinuousMultivariate}, and copula-related multivariate distributions with Gamma marginals. We propose a very general and flexible Bayesian semiparametric model specification of the multivariate MEM, with Dirichlet Process Mixture (DPM) innovation errors.

To our knowledge, the most general framework in this context has been formalized in \citet{Cipollini2013SemiparametricVMEM}, with the distribution of the error component left unspecified, except for the conditional moments. The motivation for their Generalized Method of Moments (GMM) approach relies on the limitations of complete parametric specifications of the error distribution, that is: (i) parametric distributions restricted to the positive orthant are often not flexible enough or they lack a closed-form density function; 
(ii) copulas often imply unrealistic symmetries, do not always model adequately the association among components of the error term, and require expensive tuning in cases of different marginal distributions;
(iii) the exact distribution of the error term may be not interesting, in particular when the analysis is mainly focused on the dynamics of the conditional mean.
Even if \citet{Cipollini2013SemiparametricVMEM} permits a more flexible model specification that is not forced by unhandy parametric assumptions, on the other it introduces problems related to the arbitrary choice of the first two moment conditions. Such a choice can have a crucial impact on the performance on the asymptotic approximation of a GMM estimator, especially when the degree of overidentification is large \citep{hall2005generalized}. Furthermore,  both the recursive model and the general form of vMEM in \citet{Cipollini2013SemiparametricVMEM} face non-negativity challenges, since restrictions to the positive orthant are not explicitely designed \citep{taylor2017logarithmic}.

Here we propose to flexibly assign the distribution to the error term by a Bayesian semiparametric approach, therefore not incurring in the potential problems of GMM estimation. Generalizing the approach adopted for the univariate case by \citet{Solgi2013BayesianMEM}, we then model the multidimensional distribution of the innovation vector as an infinite location-scale mixture of multidimensional kernels with support on the positive orthant. The error distribution is therefore not bounded to a special parametric form, but it is allowed to freely depart from an average parametric base model. We obtain a specification that is robust to mispecified data generating processes and that outperforms classical methods in terms of fitting and predictive power on both simulated and real data. Computational issues arising from the restrictions of the model to the positive orthant are avoided through the formulation of an efficient Markov chain monte Carlo (MCMC) algorithm on a parameter-expanded unconstrained version of the model that re-maps posterior samples to the constrained model. With proper adjustments, our proposed idea could be applied to GARCH models and to all related models for which it is easier to sample from the non-identifiable version of the model.

Approaches based on DPM have alreay been used in the econometric literature, for instance by \citet{Jensen2014Estimating} to model volatility distribution, or by \citet{Kalli2013Modeling} for financial returns. Along the same lines, \citet{jensen2010bayesian} and \citet{jensen2018risk} propose a DPM approach to stochastic volatility and financial returns, and \citet{zaharieva2020bayesian} an application to Choleski-type multivariate stochastic volatilities. 
A similar approach has been pursued in the context of multivariate GARCH models by \citet{virbickaite2016bayesian}, but in a different context: they do not have non-negativity constraints on the distribution of the innovations, explicitly ignore mean identifiability, and highlight, together with \citet{Jensen2014Estimating}, how the implementation of moment restrictions in DPM models is still an open question. 

The rest of the paper is organized as follows. In Section 2 vMEM is introduced and our semiparametric extension is presented, whilst the MCMC sampling scheme for conducting Bayesian inference is developed in Section 3. The model is evaluated in terms of fitting and predictive performance against parametric alternatives in a simulation study (Section 4) and in an empirical financial application on the interaction among volatility measures (Section 5). In the Section 6 we conclude and propose further developments.

\section{Statistical Framework for Bayesian vMEM}

Given a $d$-dimensional non-negative stochastic process $\left\{ \mathbf{x}_{t}\right\} _{t}$, the vMEM represents it as an element-by-element product of its conditional mean process $\bm{\mu}_{t}=\mathbb{E}\left[\mathbf{x}_{t}|\mathscr{F}_{t-1}\right]$, where $\mathscr{F}_{t-1}$ is the information set available at time $t-1$ (the $\sigma$-algebra generated by the history of the $d$-dimensional time series up to time $t-1$), and an innovation term $\bm{\varepsilon}_{t}$. 
Formally we will have:
\begin{equation}
\mathbf{x}_{t}=\bm{\mu}_{t}\odot\bm{\varepsilon}_{t}\label{eq:2.1}
\end{equation}
where $\odot$ is the Hadamard element-wise product and
\begin{itemize}
\item $\bm{\mu}_{t}=\bm{\mu}\left(\bm{\eta},\mathscr{F}_{t-1}\right)$ is a deterministic vector with non-negative components that evolves according to the vector of parameters $\bm{\eta}$.
\item $\bm{\varepsilon}_{t}\left|\mathscr{F}_{t-1}\right.\sim\mathscr{D}\left(\bm{\iota}_{d},\bm{\Sigma}\right)$
where $\mathscr{D}\left(\bm{\iota}_{d},\bm{\Sigma}\right)$ is a continuous
probability distribution on $\left(\mathbb{R}^{+}\right)^{d}$ with
unit-vector mean $\bm{\iota}_{d}$ and unknown constant covariance
matrix $\bm{\Sigma}$.
\end{itemize}
Thus we have that 
\begin{equation}
\begin{alignedat}{1} & E\left[\mathbf{x}_{t}\left|\mathscr{F}_{t-1}\right.\right]=\bm{\mu}_{t},\\
 & \mbox{Var}\left[\mathbf{x}_{t}\left|\mathscr{F}_{t-1}\right.\right]=\bm{\mu}_{t}\bm{\mu}_{t}^{'}\odot\bm{\Sigma}.
\end{alignedat}
\label{eq:1.2.2}
\end{equation}
and hence $\mbox{Var}\left[\mathbf{x}_{t}\left|\mathscr{F}_{t-1}\right.\right]$
is guaranteed to be a positive definite matrix. The unit mean assumption 
on the innovation term (second bullet above) is necessary to guarantee the identifiability of the model. The i.i.d.
assumption instead is not necessary: it implies that the $\mathbf{x}_{t}$s,
conditional on $\mathscr{F}_{t-1}$, are draws from a scale-family
of distributions in which the scale parameter evolves in time according
to $\bm{\mu}_{t}$ and the shape of the distribution remains unchanged,
but, in principle, as long as the conditional unit mean constraint
holds, the shape of the distribution may change through time as a
function of the elements of the information set $\mathscr{F}_{t-1}$.

\subsection{Specification of the conditional mean dynamics}

In vMEM literature, $\bm{\mu}_{t}$ is formulated as a linear combination
of the first $p$ and $q$ lagged $\mathbf{x}_{t}$\textquoteright s
and $\bm{\mu}_{t}$\textquoteright s, respectively:
\[
\bm{\mu}_{t}=\bm{\omega}+\underset{{\scriptstyle i=1}}{\overset{{\scriptstyle q}}{\sum}}\mathbf{B}_{i}\bm{\mu}_{t-i}+\underset{{\scriptstyle i=1}}{\overset{{\scriptstyle p}}{\sum}}\mathbf{A}_{i}\mathbf{x}_{t-i}
\]
With this structure, the persistence property and the interdependencies
between the components of $\mathbf{x}_{t}$ can be modelled parsimoniously.
With $p=q=1$ we obtain the base vMEM:

\begin{equation}
\bm{\mu}_{t}=\bm{\omega}+\mathbf{B}\bm{\mu}_{t-1}+\mathbf{A}\mathbf{x}_{t-1}
\label{eq:2.3-1}
\end{equation}
where

\begin{flalign*}
 & \bm{\omega}=\left[\begin{array}{c}
\omega_{1}\\
\vdots\\
\omega_{d}
\end{array}\right],\ \mathbf{B}=\left[\begin{array}{ccc}
\beta_{1,1} & \ldots & \beta_{1,d}\\
\vdots & \ddots & \vdots\\
\beta_{d,1} & \ldots & \beta_{d,d}
\end{array}\right],\ \mathbf{A}=\left[\begin{array}{ccc}
\alpha_{1,1} & \ldots & \alpha_{1,d}\\
\vdots & \ddots & \vdots\\
\alpha_{d,1} & \ldots & \alpha_{d,d}
\end{array}\right].
\end{flalign*}

Sufficient conditions so that $\bm{\mu}_{t}\in\mathbb{R}^{+^{d}}$ for all $t\geq0$
are again that all parameters $\omega_{i},\ \beta_{ij},\ \alpha_{ij}$
are positive for every $i,j=1,\ldots,d$. Yet this is not a necessary condition and, in multivariate
context, it is also quite restrictive. It is therefore often omitted
in applications in favour of a simple non-negativity check of the
values of the conditional mean obtained with the estimates of its
parameters. From the classical theory
of vector autoregressive models, sufficient conditions for stationarity of $\bm{\mu}_{t}$
are that all characteristic roots of $\mathbf{B+}\mathbf{A}$ lie
inside the unit circle. 

Several generalizations of this specification has been proposed. To
study inter-dependencies across volatility measures, \citet{Cipollini2013SemiparametricVMEM}
included an asymmetric effect depending on the returns of the underlying
index. \citet{Engle2009MEMAnalisys} and \citet{Engle2012VolatilitySpilloversInAsia},
used another augmented version of the base specification in \eqref{eq:2.3-1} including
dummies to differentiate between specific time periods to describe
the volatility spillover effect in the East Asian financial markets
before, during and after the Asian currency crisis of 1997--1998.

\subsection{Dirichlet process mixture error terms}

We model the innovation term using a mixture of simple multivariate
distributions. From a Bayesian perspective, a finite mixture with
$K$ components can be formulated as
\begin{alignat*}{1}
\bm{\varepsilon}_{t}|d_{t},\bm{\phi} & \sim F\left(\bm{\phi}_{d_{t}}\right)\\
d_{t}|\mathbf{p} & \sim\mbox{Discrete}\left(p_{1},\ldots,p_{K}\right)
\end{alignat*}
where $\bm{\phi}_{d_{t}}=\left(\phi_{1}^{\left(d_{t}\right)},\ldots,\phi_{K}^{\left(d_{t}\right)}\right)$,
$\mathbf{p}=\left(p_{1},\dots,p_{K}\right)$ and $d_{t}$ are categorical
variables (also called ``latent labels'') that determine to which
mixture component the corresponding $\bm{\varepsilon}_{t}$ belongs.
In order to fully specify this model in a Bayesian setting, we should
assign priors to $\bm{\phi}_{d}$ and $\mathbf{p}$:
\begin{alignat*}{1}
\bm{\phi}_{d} & \sim G_{0}\\
\mathbf{p} & \sim\mbox{Dir}\left(\frac{\alpha}{K},\ldots,\frac{\alpha}{K}\right)
\end{alignat*}
where $G_{0}$ is a distribution on the parameter space of $F$ and
$\mbox{Dir}\left(\frac{\alpha}{K},\ldots,\frac{\alpha}{K}\right)$
is the Dirichlet distribution on the $K$-dimensional simplex. There
are two important problems with finite component mixtures: it is usually
difficult to determine a priori the required number of components
and they lack the degree of flexibility that is needed in many applications.
To solve these problems we will use the DPM introduced by \citet{Antoniak1974Mixtures}, that can be seen
as the limit of the finite mixture specified above for $K\to\infty$. 
The stick-breaking representation of the Dirichlet Process (DP) implies that a DPM can be represented as: 
\[
\begin{alignedat}{1}f_{\bm{\varepsilon}}\left(e\right) & =\underset{{\scriptstyle j=1}}{\overset{{\scriptstyle \infty}}{\sum}}w_{j}k\left(e|\bm{\theta}_{j}\right)\\
\mathbf{w} & \sim GEM\left(\alpha\right)\\
\theta_{j} & \overset{{\scriptstyle i.i.d.}}{\sim}G_{0},
\end{alignedat}
\]
where $k$ is a parametric distribution (for our purposes constrained to the positive orthant) and $\mathbf{w}=\left\{ w_{1},w_{2},\dots\right\}$ is the the stick-breaking weight process, here represented using the notation $\mathbf{w}\sim GEM\left(\alpha\right)$ (that stands for Griffiths, Engen, McCloskey), as used in \citet{Pitman2002Combinatorial} and \citet{Johnson1997DiscreteMultivariate}. Therefore this approach bypasses the problem of choosing the correct number of components.
Furthermore, \citet{dalal1980approximating}  establish the \textit{large support} property, or adequacy, of DPM, in that a  parametric  Bayesian  model  can  be  approximated  by  a nonparametric Bayesian model with a mixture of DPs, with the prior assigning most of its weight to neighborhoods of the parametric model, and show that any parametric or nonparametric prior may  be  approximated  arbitrarily  closely  by  a  prior  which  is  a  mixture  of  DPs.   See also \citet{peluso2017robust} for  similar results on mixture processes of DPs.  Corollary 2.2 of \citet{korwar1973contributions} also supports the idea that the model under consideration has a wide applicability to various mispecifications in the generating process.

In the context of DPMs the concentration parameter $\alpha$ of the Dirichlet Process, can be interpreted as the prior belief about the number of components in the mixture: small values of $\alpha$ assume a priori an infinite mixture model with a small number of components with large weights while, large values of $\alpha$ assume a priori an infinite mixture model with all the weights being very small. As pointed out by \citet{Kalli2013Modeling}, $\alpha$ controls the exponential decay of weights and this might be a disadvantage of DPM models in the case that more mixture components are needed. An alternative could be to consider more general stick-breaking processes, for more details refer to \citet{Kalli2013Modeling}.

\subsection{Identifiability issues}
Since the data vector we consider belongs to the positive orthant
we choose to use multivariate log-normal densities as a convenient choice for the kernels of the infinite mixture.  A multivariate log-normal distribution is preferred to a multivariate Gamma, that would had been the direct multivariate generalization of the univariate approach suggested by \citet{Solgi2013BayesianMEM}, since all multivariate generalizations of the Gamma distribution we are aware of are defined via the joint characteristic function and thus require numerical inversion formulas to find the corresponding density. 

As mentioned earlier, in parametric vMEM the distribution of innovations is restricted to have unit-vector mean. If this were not true, this would directly impact the mean vector of the observations: it would be $\bm{\mu}_{t} \odot E\left[\bm{\varepsilon}\right]$  and thus any estimate of parametrs of $\bm{\mu}_{t}$ would not be interpretable. Hence, at a first glance, it seems natural to use multivariate log-normal densities with unit-vector mean and positive-definite scale matrices, $\bm{\Sigma}_{j}$. If we have a $d$-dimensional multivariate log-normal random variable $\bm{\varepsilon}$ with log-scale $\bm{m}$ and shape matrix $\bm{\Sigma}$ (i.e. we have a multidimensional random variable $\bm{\varepsilon}$ such that $\mathbf{y}=\log\bm{\varepsilon}\sim N_{d}\left(\bm{m},\bm{\Sigma}\right))$ we have that the components of the mean vector are
$
E\left[\varepsilon_{i}\right]=e^{m_{i}+\frac{1}{2}\Sigma_{i,i}}\ \forall i=1,\ldots,d,
$
and hence we obtain a unit mean vector if and only if 
$
m_{i}=-\frac{1}{2}\Sigma_{i,i}\ \forall i=1,\ldots,d,
$ therefore obtaining
\begin{equation}
\begin{alignedat}{1}f_{\bm{\varepsilon}}\left(\cdot\right) & =\underset{{\scriptstyle j=1}}{\overset{{\scriptstyle \infty}}{\sum}}w_{j}\text{logN}_{d}\left(\cdot|\mathbf{m},\bm{\Sigma}{}_{j}\right)\\
\text{logN}_{d}\left(\bm{\varepsilon}|\mathbf{m},\bm{\Sigma}\right) & =\left(\underset{{\scriptstyle i=1}}{\overset{{\scriptstyle d}}{\prod}}\frac{1}{\varepsilon_{i}}\right)\left(2\pi\right)^{-\frac{d}{2}}|\bm{\Sigma}|^{-\frac{1}{2}}\exp\left\{ -\frac{1}{2}\left(\log\bm{\varepsilon}-\mathbf{m}\right)^{'}\bm{\Sigma}^{-1}\left(\log\bm{\varepsilon}-\mathbf{m}\right)\right\} \mathbb{I}_{\left(\mathbb{R}^{+}\right)^{d}}\left(\bm{\varepsilon}\right)\\
m_{i} & =-\frac{1}{2}\Sigma_{i,i}\ \quad i=1,\ldots,d.\\
\mathbf{w} & \sim GEM\left(\alpha\right)\\
\bm{\Sigma}_{j} & \overset{{\scriptstyle i.i.d.}}{\sim}G_{0}.
\end{alignedat}
\label{3.1}
\end{equation}

Imposing all component means to only depend on the diagonal elements of the component-specific scale matrix
restricts in some way the ability of the model to cover all the possible
distributions on the positive orthant. In fact, in the univariate
log-normal case (and in the univariate Gamma case with
one parameter of \citealt{Solgi2013BayesianMEM}) the introduction of components with thicker right tails 
will increase the probability
of the neighbourhood around zero, to keep the mean fixed. Hence, in presence of fat-tailed
innovation errors, while this univariate DPM attempts to assign
higher weights to the components with smaller precision, it will,
at the same time, increases the likelihood of innovations close
to zero. In the multivariate case, the same reasoning is valid for marginals and extended to the joint distribution.
As a consequence, this model does not effectively range over all the
possibly true distributions on the positive orthant.

Then, in a more flexible view, we can replace the previous kernels
with log-normal densities with location vectors $\mathbf{m}_{j}$,
obtaining

\begin{equation}
\begin{alignedat}{1}f_{\bm{\varepsilon}}\left(\cdot\right) & =\underset{{\scriptstyle j=1}}{\overset{{\scriptstyle \infty}}{\sum}}w_{j}\text{logN}_{d}\left(\cdot|\mathbf{m}_{j},\bm{\Sigma}{}_{j}\right)\\
\text{logN}_{d}\left(\bm{\varepsilon}|\mathbf{m},\bm{\Sigma}\right) & =\left(\underset{{\scriptstyle i=1}}{\overset{{\scriptstyle d}}{\prod}}\frac{1}{\varepsilon_{i}}\right)\left(2\pi\right)^{-\frac{d}{2}}|\bm{\Sigma}|^{-\frac{1}{2}}\exp\left\{ -\frac{1}{2}\left(\log\bm{\varepsilon}-\mathbf{m}\right)^{'}\bm{\Sigma}^{-1}\left(\log\bm{\varepsilon}-\mathbf{m}\right)\right\} \mathbb{I}_{\left(\mathbb{R}^{+}\right)^{d}}\left(\bm{\varepsilon}\right)\\
\mathbf{w} & \sim GEM\left(\alpha\right)\\
\left(\mathbf{m},\bm{\Sigma}\right)_{j} & \overset{{\scriptstyle i.i.d.}}{\sim}G_{0}
\end{alignedat}
\label{3.2}
\end{equation}
By this definition clearly $f_{\bm{\varepsilon}}\left(\cdot\right)$
does not have unit mean. In fact, if we call $\bm{\sigma}_{j}=\left(\sigma_{1,1}^{\left(j\right)},\ldots,\sigma_{d,d}^{\left(j\right)}\right)$
the vector of the diagonal elements of the matrix $\bm{\Sigma}_{j}$,
we have
\[
\bar{\mathbf{m}}=\mathbb{E}_{f}\left[\bm{\varepsilon}\right]=\underset{{\scriptstyle j=1}}{\overset{{\scriptstyle \infty}}{\sum}}w_{j}\exp\left\{ \mathbf{m}_{j}+\frac{1}{2}\bm{\sigma}_{j}\right\} \neq\bm{\iota}_{d}.
\]
To solve the arising identification issue we could modify the support
of the random mixing distribution so that the infinite mixture has
unit-vector mean. This could be done simply modifying the mixture
kernels so that the density function of the innovations results specified
as
\[
g_{\bm{\varepsilon}}\left(\cdot\right)=\underset{{\scriptstyle j=1}}{\overset{{\scriptstyle \infty}}{\sum}}w_{j}\,\text{logN}_{d}\left(\cdot|\mathbf{m}_{j}-\log\bar{\mathbf{m}},\bm{\Sigma}{}_{j}\right).
\]
This specification ensures that
\[
\begin{alignedat}{1}\mathbb{E}_{g}\left[\bm{\varepsilon}\right] & =\underset{{\scriptstyle j=1}}{\overset{{\scriptstyle \infty}}{\sum}}w_{j}\exp\left\{ \mathbf{m}_{j}-\log\bar{\mathbf{m}}+\frac{1}{2}\bm{\sigma}_{j}\right\} =\\
 & =\underset{{\scriptstyle j=1}}{\overset{{\scriptstyle \infty}}{\sum}}w_{j}\exp\left\{ \mathbf{m}_{j}+\frac{1}{2}\bm{\sigma}_{j}\right\} \odot\exp\left\{ \log1/\bar{\mathbf{m}}\right\} =\\
 & =\left[\underset{{\scriptstyle j=1}}{\overset{{\scriptstyle \infty}}{\sum}}w_{j}\exp\left\{ \mathbf{m}_{j}+\frac{1}{2}\bm{\sigma}_{j}\right\} \right]\oslash\bar{\mathbf{m}}=\bm{\iota}_{d},
\end{alignedat}
\]
where $\oslash$ is the Hadamard point-wise division.
Note that the sequence of expected weights decays exponentially, with rate of decay depending on the concentration parameter $\alpha$. Furthermore, $w_j$ is stochastically independent from $(m_j,\Sigma_j^{-1} )$, whose distribution has finite mean. Therefore the mean vector decays exponentially with $j$ and thus the sum should be finite.  This also holds a posteriori, since the posterior mean vector of the Normal-Wishart depends only on the prior hyperparameters and on number of observations assigned to each component, whilst the posterior mean of the mixture weights still decays exponentially with rate depending on $\alpha+T$. Combining this model for the distribution of innovations with \eqref{eq:2.3-1}
results in a model that we will call DPMLN2-vMEM. 

\section{Parameter-Expanded Slice Sampler for Posterior Inference}

\subsection{Parameter expansion of the constrained model}
	\citet{yang2010semiparametric} introduced the idea that an unconstrained model can be seen as a parameter expansion of a constrained model and applyed it to latent factor models and, more in general, hierarchical models with latent variables.
	Following this idea, for the purpose of the sampling algorithm, we start from the unconstrained DPM for the distribution of the innovations,
which is a parameter expanded (in the sense of \citealt{Liu1999ParameterExpansion}, \citealt{VanDyk2001art}
and \citealt{Liu1998ParameterExpansionEM}) version of the DPMLN2-vMEM in \eqref{3.1} and will
be called PX-DPMLN2-vMEM. It is important to enlighten that a prior
on the parameters of the PX model induces a prior on the parameters
of the original model and that the use of proper priors results in
proper posteriors for this model (even if the likelihood is improper). 

Hence we will set up a Markov Chain Monte Carlo (MCMC) simulation
to target the PX-DPMLN2-vMEM and, at the end of this simulation, we
will post-process the sample obtained from the parameter-expanded
model to an equivalent sample from the DPMLN2-vMEM. To map the sample
from the PX-DPMLN2-VMEM to one from the DPMLN2-vMEM we will use this
transformation:
\begin{equation}
\begin{alignedat}{2}\left(\begin{array}{c}
\bm{\omega}\\
\mathbf{B}\\
\mathbf{A}\\
w_{1}\\
w_{2}\\
\vdots\\
\mathbf{m}_{1}\\
\mathbf{m}_{2}\\
\vdots\\
\bm{\Sigma}_{1}\\
\bm{\Sigma}_{2}\\
\vdots
\end{array}\right) & \rightarrow & \left(\begin{array}{c}
\bm{\omega}\odot\bar{\mathbf{m}}\\
\mathbf{B}\\
\mathbf{A}\odot\left[\bar{\mathbf{m}}\bm{\iota}_{d}^{'}\right]\\
w_{1}\\
w_{2}\\
\vdots\\
\mathbf{m}_{1}-\log\left(\bar{\mathbf{m}}\right)\\
\mathbf{m}_{2}-\log\left(\bar{\mathbf{m}}\right)\\
\vdots\\
\bm{\Sigma}_{1}\\
\bm{\Sigma}_{2}\\
\vdots
\end{array}\right)\end{alignedat}
\label{eq:3.3.4}
\end{equation}
Note that, in order to use this post-processing function, we need
to sample $\bar{\mathbf{m}}$, the mean of the DPM, that is an infinite
sum. Although the distribution of the mean of the DP and DPMs has
been the subject of several studies (for further insights see \citealt{Lijoi2004MeansOfaDP}, \citealt{Cifarelli1990Distribution}
and \citealt{Regazzini2003Distributional}), we are not aware of a simple way to sample from these distributions
since their evaluation is generally subject to computation of some
numerical integrals.

To solve this problem here we propose to approximate the infinite
sum substituting $\bar{\mathbf{m}}$ by a finite one so that the truncated
sum of weights is close enough to 1. In practice, in order to obtain
a sample from the mean of the DPM, we need to truncate $\bar{\mathbf{m}}$ at 
\begin{equation}
K_{\varepsilon_{\bar{m}}}=\inf\left\{ k\in\mathbb{N}\left|1-\underset{{\scriptstyle j=1}}{\overset{{\scriptstyle k}}{\sum}}w_{j}<\varepsilon_{\bar{m}}\right.\right\} \label{eq:3.3.5}
\end{equation}
where $\varepsilon_{\bar{m}}$ is a fixed tolerance level. In \citet{Muliere1998Approximating}
it has been shown that 
\[
K_{\varepsilon_{\bar{m}}}-1\sim\mbox{Poisson}\left(-\alpha\log\varepsilon_{\bar{m}}\right),
\]
therefore the expected value of the truncation level $K_{\varepsilon_{\bar{m}}}$
is proportional to $-\log\varepsilon_{\bar{m}}$ so that, with a small
value of the concentration parameter $\alpha$, extremely accurate
results may be obtained in a reasonable computational time.

\subsection{Slice sampler development}
Bayesian Inference on DPM models has the big issue that DPMs are
infinite dimensional objects. There are substantially two main families
of methods to deal with this problem: the ``marginal methods'',
that are based on integrating out the random distribution and the
``conditional methods''\textbf{ }that explicitly instantiate the
DP and rely on its stick-breaking representation. One of the most
used conditional methods is the so-called ``slice sampler'', introduced
by \citet{Walker2007Sampling}. Here we will describe, adapt to our
goals and finally use its efficient version due to \citet{Kalli2011EfficientSliceSampler}, that in the original paper had been used to sample from DPMs but also from other mixtures based on normalized weights.

Following \citet{Walker2007Sampling}, we augment the model with the latent variable $u$ such that the joint density of $\left(\bm{\varepsilon},u\right)$ is 
\begin{equation}
f_{\bm{\varepsilon},u}\left(\bm{\varepsilon},u\right)=\underset{{\scriptstyle j=1}}{\overset{{\scriptstyle \infty}}{\sum}}\mathbb{I}\left(w_{j}>u\right)\text{logN}_{d}\left(\bm{\varepsilon}\left|\mathbf{m}_{j},\bm{\Sigma}_{j}\right.\right),\label{4.1}
\end{equation}
Therefore, given $u$, the infinite mixture reduces to a finite mixture:
for every fixed value of $u$ in $\left[0,1\right]$, only a finite
number of $w_{j}s$ can be greater than $u$, since $\underset{{\scriptstyle j=1}}{\overset{{\scriptstyle \infty}}{\sum}}w_{j}=1$.
Moreover, introducing the latent label $l$ that indicates to which
component of the mixture $\bm{\varepsilon}$ belongs, the joint density
of $\left(\bm{\varepsilon},u,l\right)$ is
\begin{equation}
f_{\bm{\varepsilon},u,l}\left(\bm{\varepsilon},u,l\right)=\mathbb{I}\left(w_{l}>u\right)\text{logN}_{d}\left(\bm{\varepsilon}\left|\mathbf{m}_{l},\bm{\Sigma}_{l}\right.\right).\label{4.2}
\end{equation}
Obviously it is not possible to sample the infinite set of parameters
$\left(\mathbf{m}_{j},\bm{\Sigma}_{j}\right)_{j>1}$ but it had been
shown by \citet{Walker2007Sampling} that, by augmenting the model
with the latent variable $u$, we only need to sample a finite set
of these parameters to obtain a sample from the target ``DPM distribution''
(i.e. distribution that is a trajectory of a DPM). 

In order to improve the efficiency of the slice sampler, \citep{Kalli2011EfficientSliceSampler}
proposed to sample in a block $u$ and $w$ and to rewrite the joint
density \eqref{4.2} as
\[
f_{\bm{\varepsilon},u,l}\left(\bm{\varepsilon},u,l\right)=\mathbb{I}\left(\xi_{l}>u\right)\frac{w_{l}}{\xi_{l}}\text{logN}_{d}\left(\bm{\varepsilon}\left|\mathbf{m}_{l},\bm{\Sigma}_{l}\right.\right),
\]
where $\left\{ \xi_{l}\right\} $ is an infinite sequence decreasing
in $l$. The block sampling increases the efficiency with respect
to the original algorithm since $u$ and $w$ are strongly correlated,
while the introduction of $\xi_{l}$ reduces the sampling of useless
$w_{j}s$. 
In what follows, we will use a deterministic, decreasing sequence
$\left\{ \xi_{j}\right\} _{j\in\mathbb{N}}$ but, in general, a random
sequence could also be considered. \citet{Kalli2011EfficientSliceSampler}
found that the mixing of the resulting Markov chain depends on the
rate of increase of $\frac{E\left[w_{j}\right]}{\xi_{j}}$: higher
rates of increase are associated with better mixing but longer running
times, since the average size of the sets $\left\{ j\left|w_{j}>u\right.\right\} $
increases. They suggest increasing the rate of increase of $\frac{E\left[w_{j}\right]}{\xi_{j}}$
until the gains in mixing are counter-balanced by the longer running
time. In their examples, \citet{Kalli2011EfficientSliceSampler} find
that $\frac{E\left[w_{j}\right]}{\xi_{j}}\propto\left(\frac{3}{2}\right)^{j}$
strikes a good balance. Thus we set $\xi_{j}\propto\frac{E\left[w_{j}\right]}{\left(1.5\right)^{j}}$. We have $\mathbf{x}_{t}\oslash\bm{\mu}_{t}=\bm{\varepsilon}_{t}$ and $f_{\bm{\varepsilon}_{t}}\left(\cdot\right)=\underset{{\scriptstyle j=1}}{\overset{{\scriptstyle \infty}}{\sum}}w_{j}\text{logN}_{d}\left(\cdot\left|\mathbf{m}_{j},\bm{\Sigma}{}_{j}\right.\right)$,
then
\small
\[
\begin{aligned}f_{\mathbf{x}_{t}|\ldots}\left(\bm{x}\right) & =f_{\bm{\varepsilon}_{t}}\left(\bm{x}\oslash\bm{\mu}\right)\left|\frac{\partial\bm{\varepsilon}}{\partial\bm{x}}\right|=\\
 & =\underset{{\scriptstyle j=1}}{\overset{{\scriptstyle \infty}}{\sum}}w_{j}\text{logN}_{d}\left(\bm{x}\oslash\bm{\mu}\left|\mathbf{m}_{j},\bm{\Sigma}_{j}\right.\right)\overset{{\scriptstyle d}}{\underset{{\scriptstyle i=1}}{\prod}}\frac{1}{\mu_{i}}=\\
 & =\underset{{\scriptstyle j=1}}{\overset{{\scriptstyle \infty}}{\sum}}w_{j}\overset{{\scriptstyle d}}{\underset{{\scriptstyle i=1}}{\prod}}\frac{1}{\mu_{i}}\overset{{\scriptstyle d}}{\underset{{\scriptstyle i=1}}{\prod}}\frac{\mu_{i}}{x_{i}}\left(2\pi\right)^{-\frac{d}{2}}\left|\bm{\Sigma}_{j}\right|{}^{-\frac{1}{2}}\exp\left\{ -\frac{1}{2}\left(\log\left(\bm{x}\oslash\bm{\mu}\right)-\mathbf{m}_{j}\right)^{'}\bm{\Sigma}^{-1}\left(\log\left(\bm{x}\oslash\bm{\mu}\right)-\mathbf{m}_{j}\right)\right\} =\\
 & =\underset{{\scriptstyle j=1}}{\overset{{\scriptstyle \infty}}{\sum}}w_{j}\overset{{\scriptstyle d}}{\underset{{\scriptstyle i=1}}{\prod}}\frac{1}{x_{i}}\left(2\pi\right)^{-\frac{d}{2}}\left|\bm{\Sigma}_{j}\right|{}^{-\frac{1}{2}}\exp\left\{ -\frac{1}{2}\left(\log\bm{x}-\log\bm{\mu}-\mathbf{m}_{j}\right)^{'}\bm{\Sigma}_{j}^{-1}\left(\log\bm{x}-\log\bm{\mu}-\mathbf{m}_{j}\right)\right\} =\\
 & =\underset{{\scriptstyle j=1}}{\overset{{\scriptstyle \infty}}{\sum}}w_{j}\text{logN}_{d}\left(\bm{x}\left|\mathbf{m}_{j}+\log\bm{\mu},\bm{\Sigma}_{j}\right.\right)
\end{aligned}
\]
\normalsize
Hence the posterior of our PX-DPMLN2-vMEM model is:
\[
\begin{aligned} & p\left(\bm{\eta},\mathbf{m}_{1},\mathbf{m}_{2},\ldots,\bm{\Sigma}_{1},\bm{\Sigma}_{2},\ldots,\mathbf{w},\mathbf{d},\mathbf{u}\left|\mathbf{x}_{1},\ldots,\mathbf{x}_{t}\right.\right)=\\
 & =\mbox{Priors}\times\underset{{\scriptstyle t=1}}{\overset{{\scriptstyle T}}{\prod}}\mathbb{I}_{\left(\xi_{d_{t}}>u_{t}\right)}\frac{w_{l_{t}}}{\xi_{l_{t}}}\,\text{logN}_{d}\left(\mathbf{x}_{t}\oslash\bm{\mu}_{t}\left|\mathbf{m}_{l_{t}},\bm{\Sigma}{}_{l{}_{t}}\right.\right)\overset{{\scriptstyle d}}{\underset{{\scriptstyle i=1}}{\prod}}\frac{1}{\mu_{i}^{\left(t\right)}}=\\
 & =\mbox{Priors}\times\underset{{\scriptstyle t=1}}{\overset{{\scriptstyle T}}{\prod}}\mathbb{I}_{\left(\xi_{l_{t}}>u_{t}\right)}\frac{w_{l_{t}}}{\xi_{l_{t}}}\overset{{\scriptstyle d}}{\underset{{\scriptstyle i=1}}{\prod}}\frac{1}{x_{i}^{\left(t\right)}}\left(2\pi\right)^{-\frac{d}{2}}\left|\bm{\Sigma}_{l_{t}}\right|{}^{-\frac{1}{2}} \cdot\\
 & \ \ \ \ \ \ \ \ \ \ \ \ \ \ \cdot \exp\left\{ -\frac{1}{2}\left(\log\mathbf{x}_{t}-\log\bm{\mu}_{t}-\mathbf{m}_{l_{t}}\right)^{'}\bm{\Sigma}_{l_{t}}^{-1}\left(\log\mathbf{x}_{t}-\log\bm{\mu}_{t}-\mathbf{m}_{l_{t}}\right)\right\} ,
\end{aligned}
\]
where $\bm{\eta}$ is the vector of all the parameters from which depends
the conditional mean, $\mathbf{w}=\left\{ w_{1},w_{2},...\right\} $
is the weight process, $\mathbf{l}=\left(l_{1},\ldots,l_{T}\right)$
is the vector of latent labels and $\mathbf{u}=\left(u_{1},\ldots,u_{T}\right)$
is the vector of latent variables such that \eqref{4.1} holds. 

In our MCMC simulations we sample $l_{t}$,$u_{t}$ $\forall t=1,\ldots,T$,
$v_{j},\mathbf{m}_{j},\bm{\Sigma}_{j}$ for all the required $j$s
and $\bm{\eta}$. Then we post process the sample obtained using the
map \eqref{eq:3.3.4} in order to obtain a sample from the posterior
of DPMLN2-vMEM. We will now detail the steps of the slice sampler.

\subsubsection{Sampling $u_{t}$}

The full conditional probability density function of $u_{t}$ is 
\[
p\left(u_{t}\left|\ldots\right.\right)\propto\mathbb{I}\left(\xi_{l_{t}}>u_{t}\right).
\]
Therefore, we can sample $u_{t}$ from the uniform distribution
$U\left(0,\xi_{l_{t}}\right).$

\subsubsection{Sampling $v_{j}$}

As described in the first chapter, $v_{j}\overset{i.i.d.}{\sim}\mbox{Beta}\left(1,\alpha\right)$.
Thus the full conditional probability density function of $v_{j}$
is 
\[
\begin{aligned}p\left(v_{j}\left|\ldots\right.\right) & \propto\pi\left(v_{j}\right)\underset{{\scriptstyle t:\,l_{t}\geq j}}{\prod}w_{l_{t}}\propto\\
 & \propto v_{j}^{0}\left(1-v_{j}\right)^{\alpha-1}\underset{{\scriptstyle t:\,l_{t}\geq j}}{\prod}v_{l_{t}}\underset{{\scriptstyle k=1}}{\overset{{\scriptstyle l_{t}-1}}{\prod}}\left(1-v_{k}\right)\propto\\
 & \propto\left(1-v_{j}\right)^{\alpha-1}\underset{{\scriptstyle t:\,l_{t}=j}}{\prod}v_{l_{t}}\underset{{\scriptstyle k=1}}{\overset{{\scriptstyle l_{t}-1}}{\prod}}\left(1-v_{k}\right)\underset{{\scriptstyle t:\,l_{t}>j}}{\prod}v_{l_{t}}\underset{{\scriptstyle k=1}}{\overset{{\scriptstyle l_{t}-1}}{\prod}}\left(1-v_{k}\right)\propto\\
 & \propto\left(1-v_{j}\right)^{\alpha-1}\underset{{\scriptstyle t:\,l_{t}=j}}{\prod}v_{l_{t}}\underset{{\scriptstyle t:\,l_{t}>j}}{\prod}\left(1-v_{j}\right)=\\
 & =v_{j}^{n_{j}}\left(1-v_{j}\right)^{\alpha-1+g_{j}}.
\end{aligned}
\]
Therefore, the full conditional distribution of $v_{j}$ is $\mbox{Beta}\left(1+n_{j},\alpha+g_{j}\right),$
where $n_{j}=\underset{t=1}{\overset{T}{\sum}}\mathbb{I}_{\left(l_{t}=j\right)}$
and $g_{j}=\underset{t=1}{\overset{T}{\sum}}\mathbb{I}_{\left(l_{t}>j\right)}$
.

Note that $\underset{t=1}{\overset{T}{\sum}}\mathbb{I}_{\left(l_{t}=j\right)}=\underset{t=1}{\overset{T}{\sum}}\mathbb{I}_{\left(l_{t}>j\right)}=0$
$\forall j\geq\bar{d}=\max\left\{ l_{1},\ldots,l_{T}\right\} $: this
means that the distribution of $v_{j}$ will be updated if and only
if there exists at least one innovation coming from a component with
index greater than (or equal to) $j$. Otherwise the full conditional
of $v_{j}$ is equal to the prior distribution. Therefore at this
step of the sampling we only need to update a finite number, $N$,
of $v_{j}$s: the others will not be updated and, if we will ever
need them in other steps of the sampler, we will sample them from
their prior.

\subsubsection{Sampling $\left(\mathbf{m}_{j},\Sigma_{j}^{-1}\right)$}

In our PX-DPMLN2-vMEM model we put a $d$-dimensional Normal-Wishart
prior on $\left(\mathbf{m}_{j},\bm{\Sigma}_{j}^{-1}\right)$. This
prior is the conjugate prior for a Bayesian model with normal data,
so we consider a transformation of the data: 
\begin{equation}
\left\{ \begin{alignedat}{1}\bm{\varepsilon}_{t}= & \mathbf{x}_{t}\oslash\bm{\mu}_{t}\ \forall t=1,\ldots,T\\
f_{\bm{\varepsilon}}\left(\cdot\right) & =\underset{{\scriptstyle j=1}}{\overset{{\scriptstyle \infty}}{\sum}}w_{j}\log N_{d}\left(\bm{\varepsilon}|\mathbf{m}_{j},\bm{\Sigma}{}_{j}\right)\\
f_{\mathbf{x}}\left(\bm{x}\right) & =f_{\bm{\varepsilon}}\left(\bm{x}\oslash\bm{\mu}\right)\left|\frac{\partial\bm{\varepsilon}}{\partial\bm{x}}\right|
\end{alignedat}
\right.\Longrightarrow\left\{ \begin{alignedat}{1}\log\bm{\varepsilon}_{t}= & \log\left(\mathbf{x}_{t}\oslash\bm{\mu}_{t}\right)=\mathbf{y}_{t}\ \forall t=1,\ldots,T\\
f_{\log\bm{\varepsilon}}\left(\cdot\right) & =\underset{{\scriptstyle j=1}}{\overset{{\scriptstyle \infty}}{\sum}}w_{j}N_{d}\left(\log\bm{\varepsilon}|\mathbf{m}_{j},\bm{\Sigma}{}_{j}\right)\\
f_{\mathbf{y}}\left(\bm{y}\right) & =f_{\log\bm{\varepsilon}}\left(\log\left(\mathbf{x}\oslash\bm{\mu}\right)\right)\left|\frac{\partial\log\bm{\varepsilon}}{\partial\log\left(\mathbf{x}\oslash\bm{\mu}\right)}\right|
\end{alignedat}
\right.\label{4.3}
\end{equation}
So for every $j=1,2,\ldots$, we put
\begin{equation}
\begin{alignedat}{1}\bm{\Sigma}_{j}^{-1} & \sim\mbox{Wishart}_{d}\left(a,\mathbf{W}\right)\\
\mathbf{m}_{j}\left|\bm{\Sigma}_{j}^{-1}\right. & \sim N_{d}\left(\bm{\nu},n_{0}\bm{\Sigma}_{j}^{-1}\right)
\end{alignedat}
\label{eq:3.4.4}
\end{equation}
where $a\geq d$, $n_{0}>0$, $\mathbf{W}$ is a positive definite,
symmetric $d\times d$ matrix and $n_{0}\bm{\Sigma}^{-1}$ is the
precision matrix. Thus we obtain that
\[
\begin{alignedat}{1} & \bm{\Sigma}_{j}^{-1}\left|\mathbf{y}_{1},\ldots,\mathbf{y}_{T}\right.\sim\\
 & \sim\mbox{Wishart}_{d}\left(a+n_{j},\,\left[\mathbf{W}^{-1}+\underset{{\scriptstyle t:\,l_{t}=j}}{\sum}\left(\mathbf{y}_{t}-\bar{\mathbf{y}}_{j}\right)\left(\mathbf{y}_{t}-\bar{\mathbf{y}}_{j}\right)^{'}+\frac{n_{0}n_{j}}{n_{j}+n_{0}}\left(\bar{\mathbf{y}}_{j}-\bm{\nu}\right)\left(\bar{\mathbf{y}}_{j}-\bm{\nu}\right)^{'}\right]^{-1}\right)\\
 & \mathbf{m}_{j}\left|\bm{\Sigma}_{j}^{-1},\mathbf{y}_{1},\ldots,\mathbf{y}_{T}\right.\sim N_{d}\left(\frac{n_{0}\bm{\nu}+n_{j}\bar{\mathbf{y}}_{j}}{n_{0}+n_{j}},\,\left(n_{0}+n_{j}\right)^{-1}\bm{\Sigma}_{j}\right)
\end{alignedat}
\]
with $\bar{\mathbf{y}}_{j}=\frac{1}{n_{j}}\underset{t:\,l_{t}=j}{\sum}\mathbf{y}_{t}$
and where $\left(n_{0}+n_{j}\right)\bm{\Sigma}_{j}^{-1}$ is the precision
matrix. Note that, although $j=1,2,\ldots$ , only a finite number
of $\left(\mathbf{m}_{j},\bm{\Sigma}_{j}^{-1}\right)s$ will be updated
at each step of the Gibbs sampler, since the full conditional of all
the couples for which $n_{j}=0$ is equal to their prior.

\subsubsection{Sampling $l_{t}$}

The full conditional distribution of $l_{t}$ is given by the probabilities:
\[
\begin{alignedat}{1} & \Pr\left\{ l_{t}=k\left|\ldots\right.\right\} \propto\\
 & \propto\mathbb{I}_{\left(\xi_{k}>u_{t}\right)}\frac{w_{k}}{\xi_{k}}\overset{{\scriptstyle d}}{\underset{{\scriptstyle i=1}}{\prod}}\frac{1}{x_{i}^{\left(t\right)}}\left(2\pi\right)^{-\frac{d}{2}}\left|\bm{\Sigma}_{k}\right|{}^{-\frac{1}{2}}\exp\left\{ -\frac{1}{2}\left(\log\mathbf{x}_{t}-\log\bm{\mu}_{t}-\mathbf{m}_{k}\right)^{'}\bm{\Sigma}_{k}^{-1}\left(\log\mathbf{x}_{t}-\log\bm{\mu}_{t}-\mathbf{m}_{k}\right)\right\} .
\end{alignedat}
\]
Since $\xi_{k}\propto\left(\frac{2}{3}\right)^{k}E\left[w_{k}\right]=\left(\frac{2}{3}\right)^{k}\frac{1}{1+\alpha}\left(\frac{\alpha}{1+\alpha}\right)^{k-1}=\frac{1}{\alpha}\left(\frac{2\alpha}{3+3\alpha}\right)^{k}$
is a decreasing function of $k$, for every $k\geq\log_{\frac{2\alpha}{3+3\alpha}}\left(\alpha u_{t}\right)$
we have that $\xi_{k}\leq u_{t}$ and hence $\Pr\left\{ l_{t}=k\left|\ldots\right.\right\} =0$.
Consequently, given all the other parameters, $l_{t}$ takes values
in the finite set $\left\{ 1,\ldots,\left\lfloor \log_{\frac{2\alpha}{3+3\alpha}}\left(\alpha u_{t}\right)\right\rfloor \right\} ,$where
$\left\lfloor a\right\rfloor $ stands for the integer part of the
real number $a$.

\subsubsection{\label{sub:Sampling eta}Sampling $\eta$}

The full conditional probability density function of the vector of
parameters of the conditional mean, $\bm{\eta}_{1\times m}=\mbox{vec}\left(\left[\bm{\omega},\mathbf{B},\mathbf{A}\right]\right)_{1\times m}$,
is
\begin{equation}
\begin{aligned}p\left(\bm{\eta}\left|\ldots\right.\right) & \propto p\left(\bm{\eta}\right)\times\underset{{\scriptstyle t=1}}{\overset{{\scriptstyle T}}{\prod}}\text{logN}_{d}\left(\mathbf{x}_{t}\oslash\bm{\mu}_{t}\left(\bm{\eta}\right)\left|\mathbf{m}_{l_{t}},\bm{\Sigma}{}_{l{}_{t}}\right.\right)\overset{{\scriptstyle d}}{\underset{{\scriptstyle i=1}}{\prod}}\frac{1}{\mu_{i}^{\left(t\right)}\left(\bm{\eta}\right)}\propto\\
 & \propto p\left(\bm{\eta}\right)\times\underset{{\scriptstyle t=1}}{\overset{{\scriptstyle T}}{\prod}}\exp\left\{ -\frac{1}{2}\left(\log\mathbf{x}_{t}-\log\bm{\mu}_{t}\left(\bm{\eta}\right)-\mathbf{m}_{l_{t}}\right)^{'}\bm{\Sigma}_{l_{t}}^{-1}\left(\log\mathbf{x}_{t}-\log\bm{\mu}_{t}\left(\bm{\eta}\right)-\mathbf{m}_{l_{t}}\right)\right\} ,
\end{aligned}
\label{eq:3.4.5}
\end{equation}
which is not a standard distribution. For the prior of $\bm{\eta}$
we use an independent Normal distribution with large variances:
\[
p\left(\bm{\eta}\right)=N_{m}\left(\bm{\eta};\mathbf{0}_{m},20\mathbf{I}_{m}\right),
\]
where $m=d+4d^{2}+d\left(\ell-1\right)+2\left(\ell-1\right)d^{2}$
and $N_{m}\left(\cdot;\mathbf{0}_{m},20\mathbf{I}_{m}\right)$ is
the density function of the $m$-dimensional Normal distribution with
parameters $\left(\mathbf{0}_{m},20\mathbf{I}_{m}\right)$.

To sample $\bm{\eta}$ we will use an adaptive version of the random-walk
Metropolis-Hastings algorithm with proposal density
\[
q\left(\bm{\eta}_{n},\bm{\eta}_{n+1}\right)=p\cdot N_{m}\left(\bm{\eta}_{n+1};\bm{\eta}_{n},\,\frac{\bm{\Lambda}_{n}}{m}\sigma_{1}^{2}\right)+\left(1-p\right)\cdot N_{m}\left(\bm{\eta}_{n+1};\bm{\eta}_{n},\,\frac{\bm{\Lambda}_{n}}{m}\sigma_{2}^{2}\right).
\]
The $\bm{\Lambda}_{n}$ component of the proposal covariance matrix
is adapted as 
\[
\bm{\Lambda}_{n}=\hat{\bm{\Sigma}}_{n}\oslash\mathbf{C}+10^{-6}\mathbf{I}_{m}
\]
where $\hat{\bm{\Sigma}}_{n}$ is the empirical covariance matrix
of the vectors obtained from $\bm{\eta}_{1},\ldots,\bm{\eta}_{n}$
using transformation \eqref{eq:3.3.4}, and 

\[
\mathbf{C}=\left[\begin{array}{c}
\bar{\mathbf{m}}\\
\bm{\iota}_{d^{2}}\\
\mathbf{\bar{m}}\\
\vdots\\
\mathbf{\bar{m}}
\end{array}\right]\left[\begin{array}{ccccc}
\mathbf{\bar{m}}' & \bm{\iota}'_{d^{2}} & \mathbf{\bar{m}}' & \ldots & \mathbf{\bar{m}}'\end{array}\right]
\]
is the transformation matrix to be applied to $\hat{\bm{\Sigma}}_{n}$
to recover from it the empirical covariance matrix of $\left[\bm{\eta}_{1},\ldots,\bm{\eta}_{k}\right]$.

For what it takes the scale parameters, $\sigma_{1}$ and $\sigma_{2}$,
and the mixture weight, $p$, we consider them as constants that should
be tuned.

Finally it is important to enlighten that at the $k$-th iteration
$\hat{\bm{\Sigma}}_{n}$ changes only by $O\left(\frac{1}{n}\right)$
by definition. Therefore this adaptation mechanism satisfies the ``diminishing
adaptation'' condition of \citep{Roberts2007Coupling} and thus the
correct target distribution is preserved.

\section{Simulation study}

\begin{figure}[H]
\caption{\label{fig:4.2.5}MCMC traces, posterior densities and ACF of the
components of the post-processed vector $\bm{\omega}$. The green
lines in the histogram represent the 95\% C.I. while the red one is
the true value.}

\centering{}\includegraphics[scale=0.12]{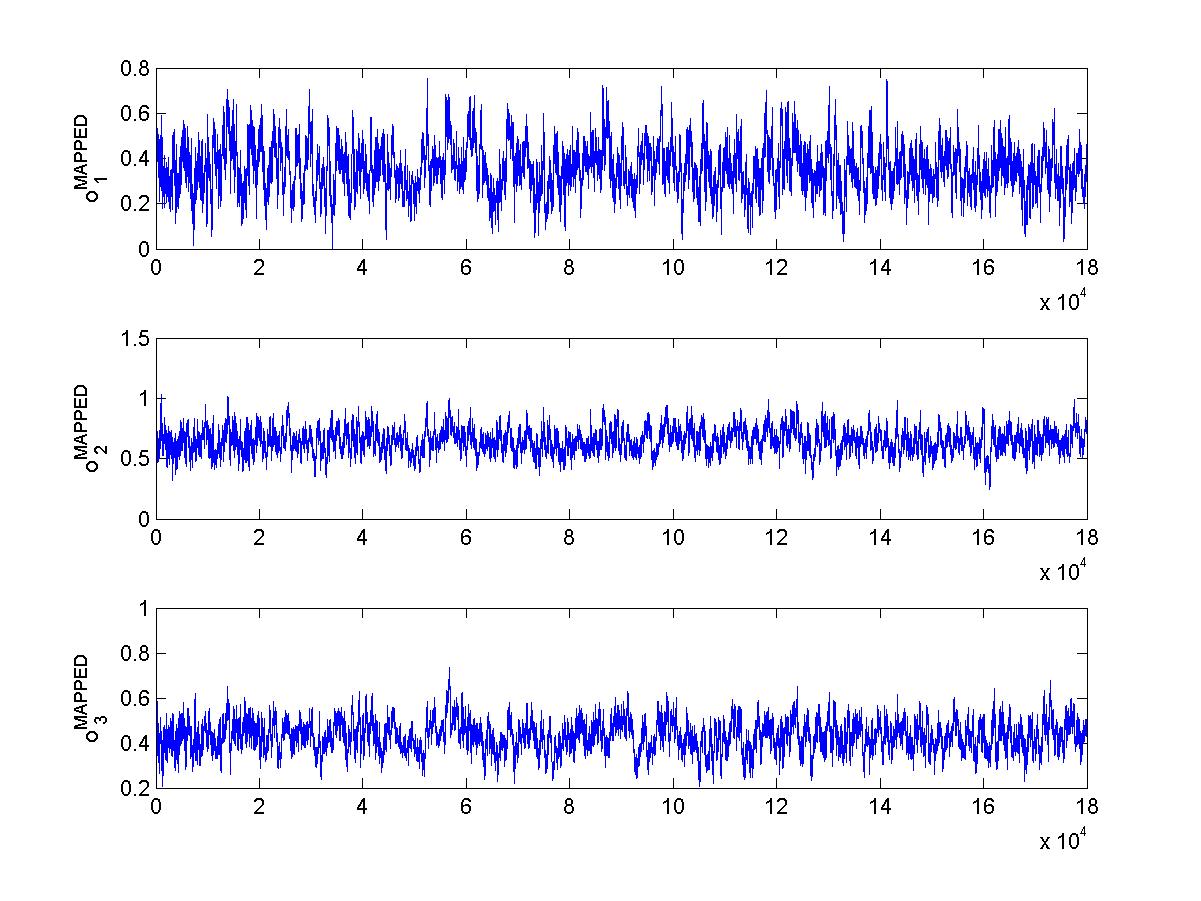}\includegraphics[scale=0.12]{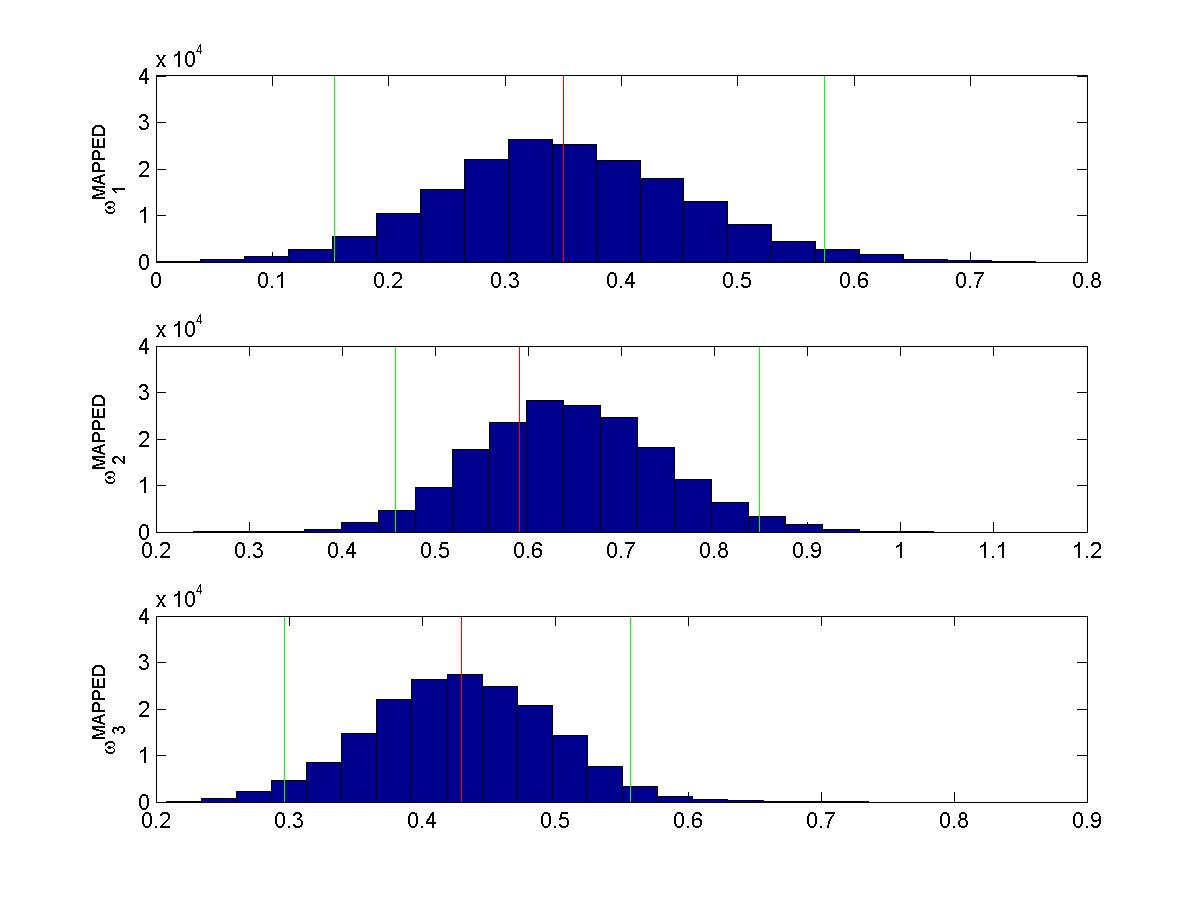}\includegraphics[scale=0.12]{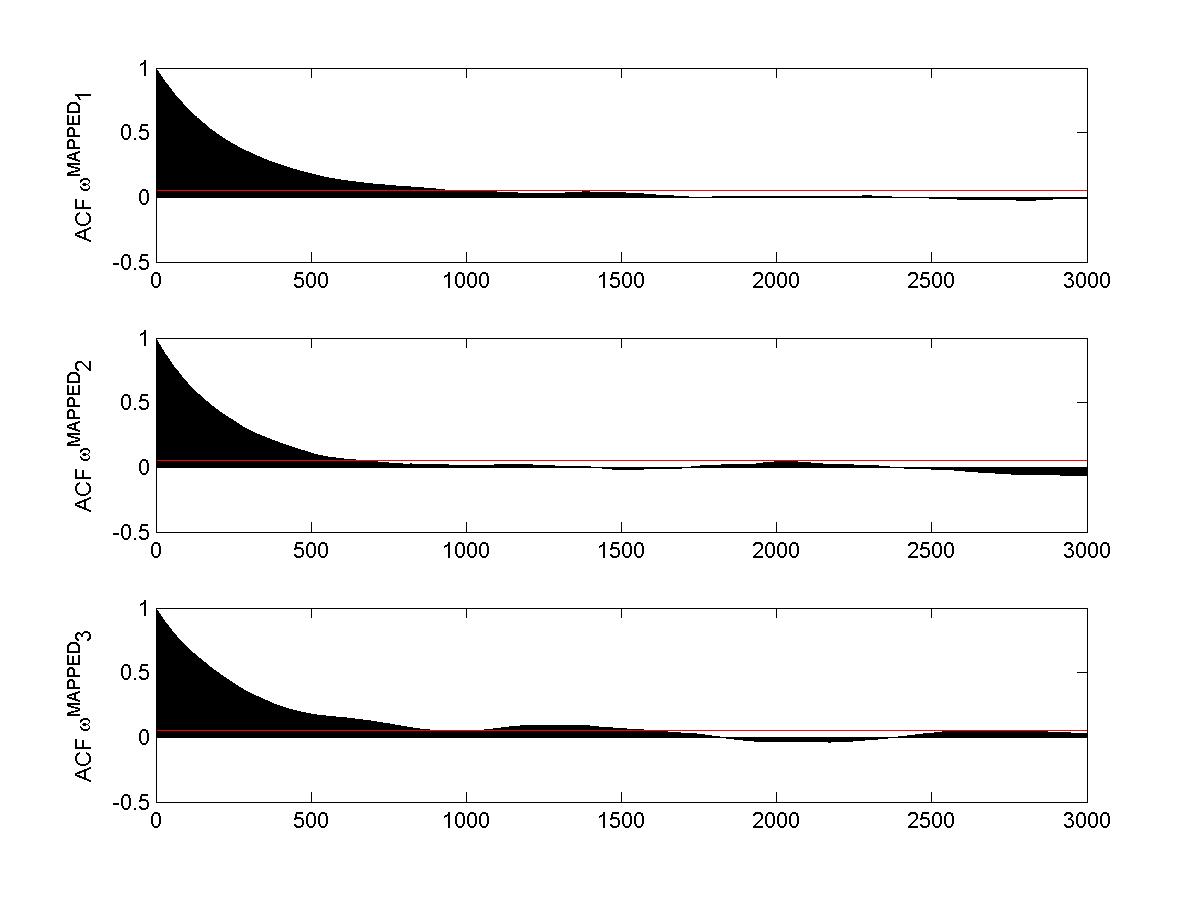}
\end{figure}

\begin{figure}[h]
\caption{\label{fig:4.2.6}MCMC traces, posterior densities and ACF of the
components of the post-processed vector $\bm{\beta}$. The green lines
in the histogram represent the 95\% C.I. while the red one is the
true value.}

\centering{}\includegraphics[scale=0.12]{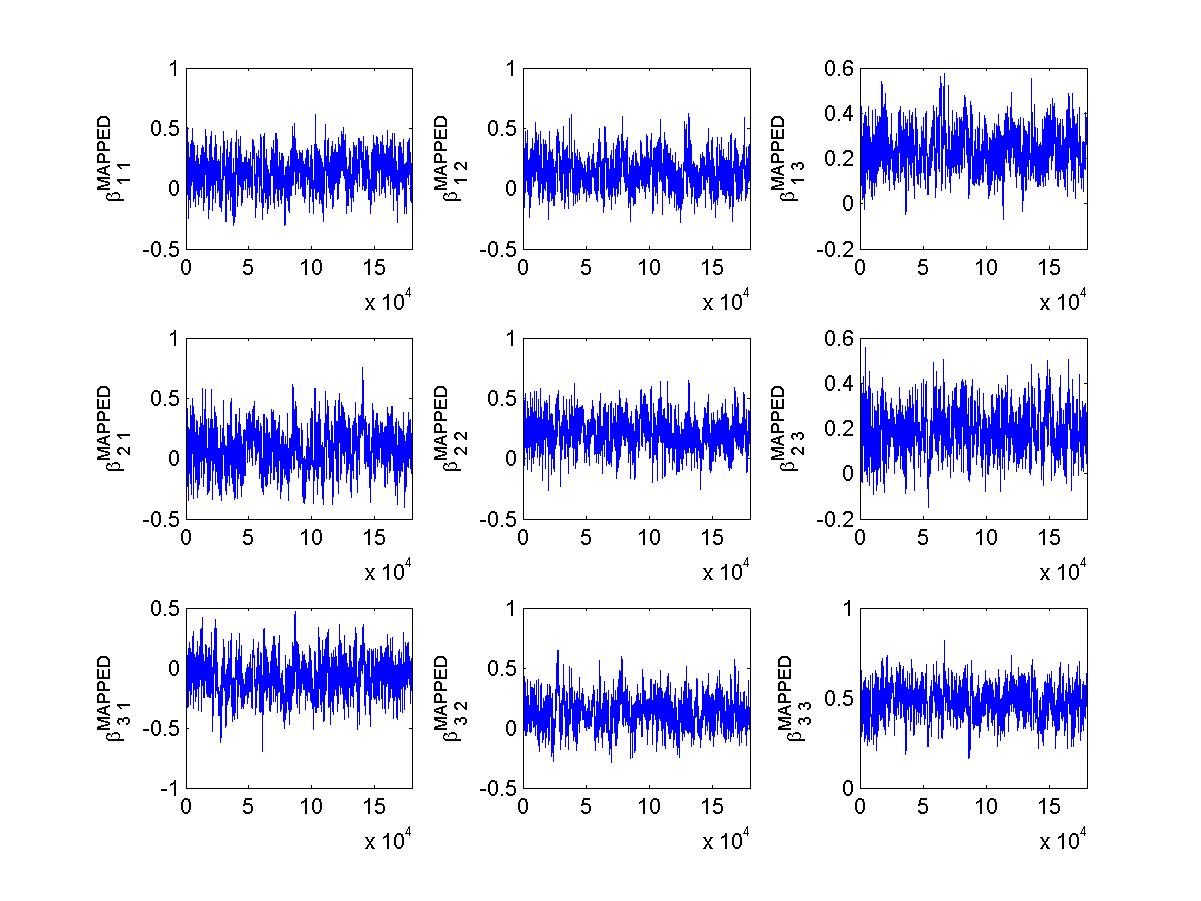}\includegraphics[scale=0.12]{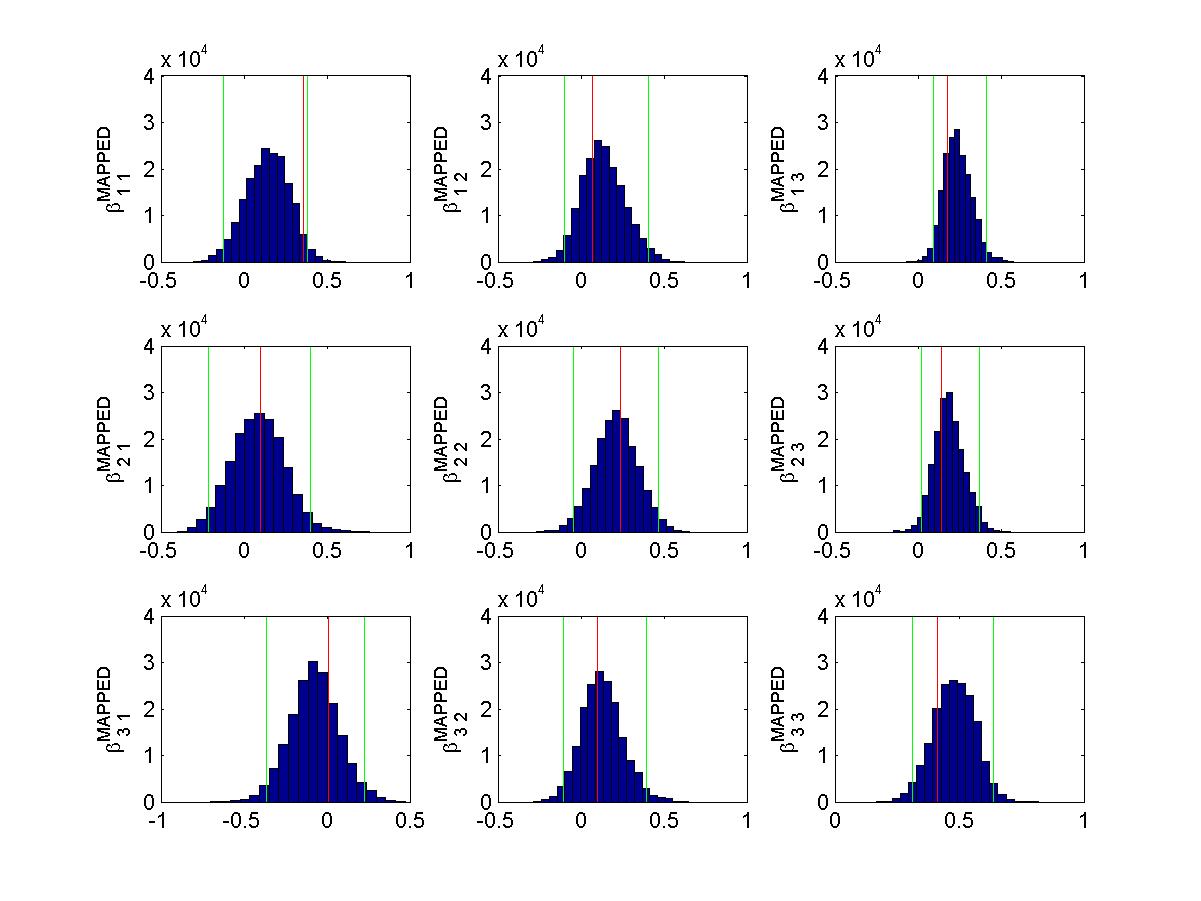}\includegraphics[scale=0.12]{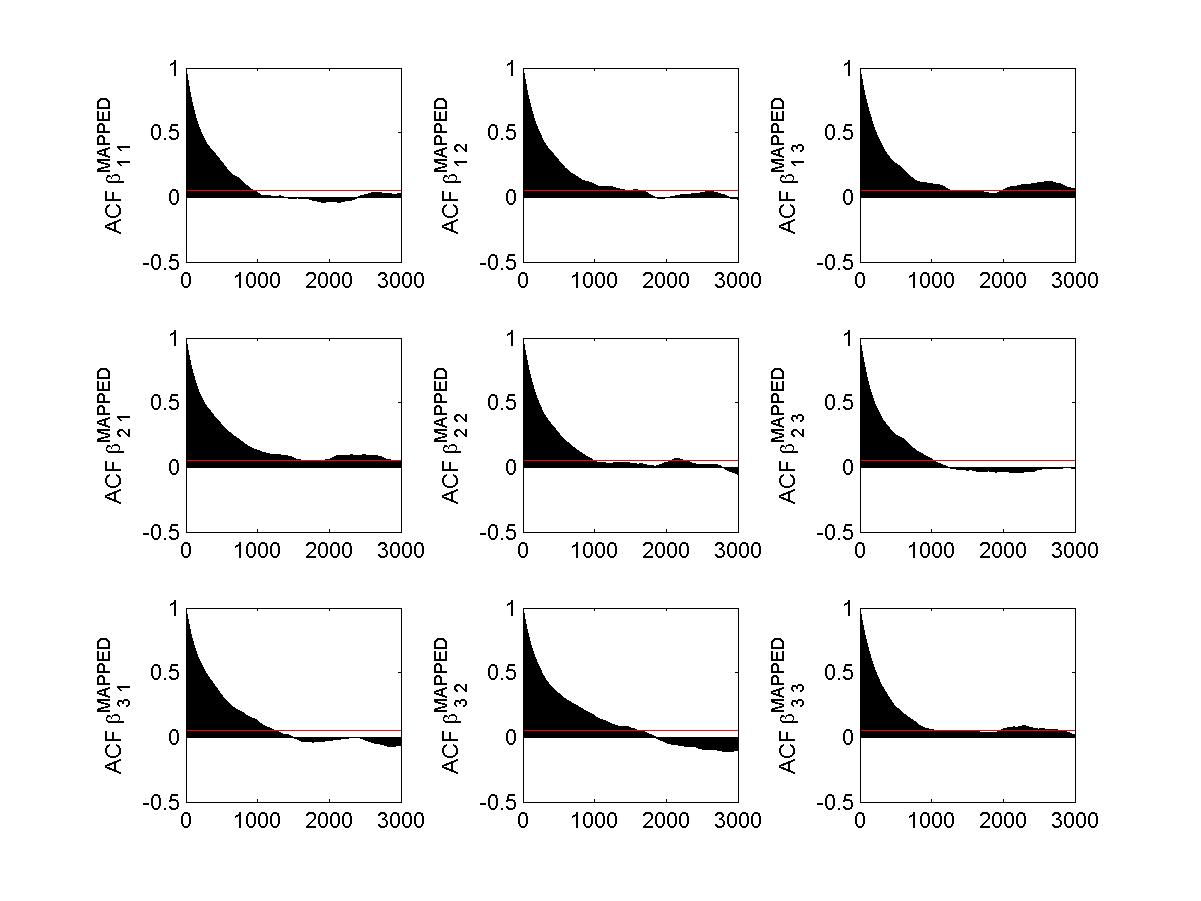}
\end{figure}

\begin{figure}[H]
\caption{\label{fig:4.2.7}MCMC traces, posterior densities and ACF of the
components of the post-processed matrix $\mathbf{A}$. The green lines
in the histogram represent the 95\% C.I. while the red one is the
true value.}

\centering{}\includegraphics[scale=0.12]{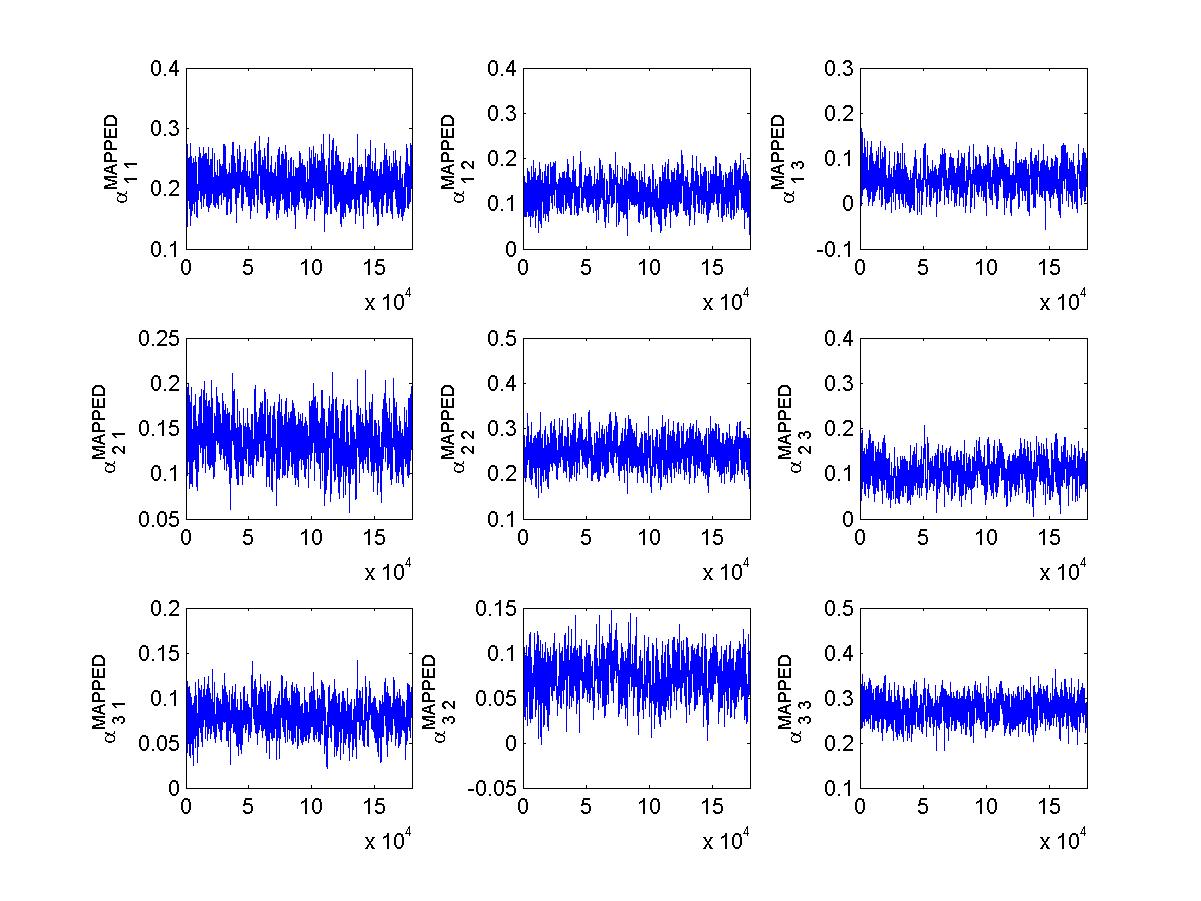}\includegraphics[scale=0.12]{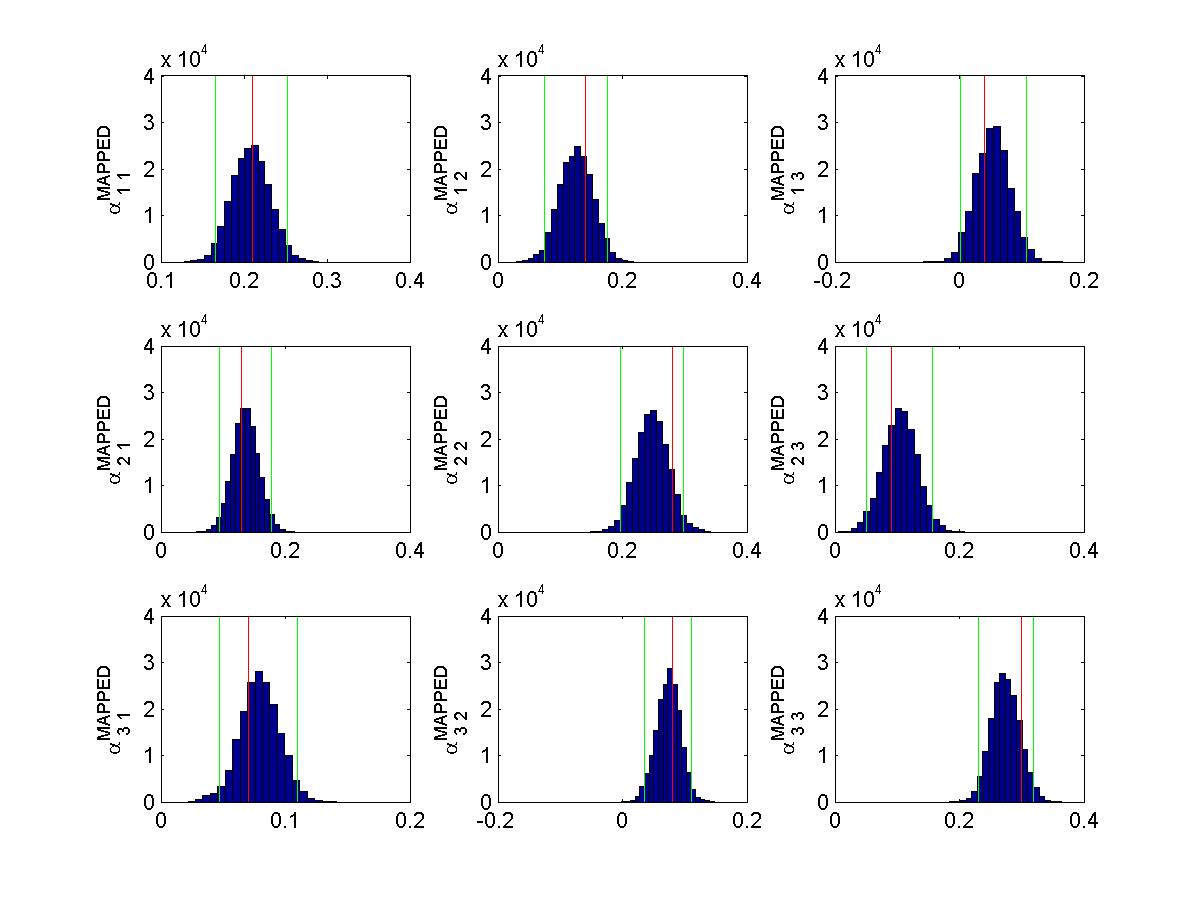}\includegraphics[scale=0.12]{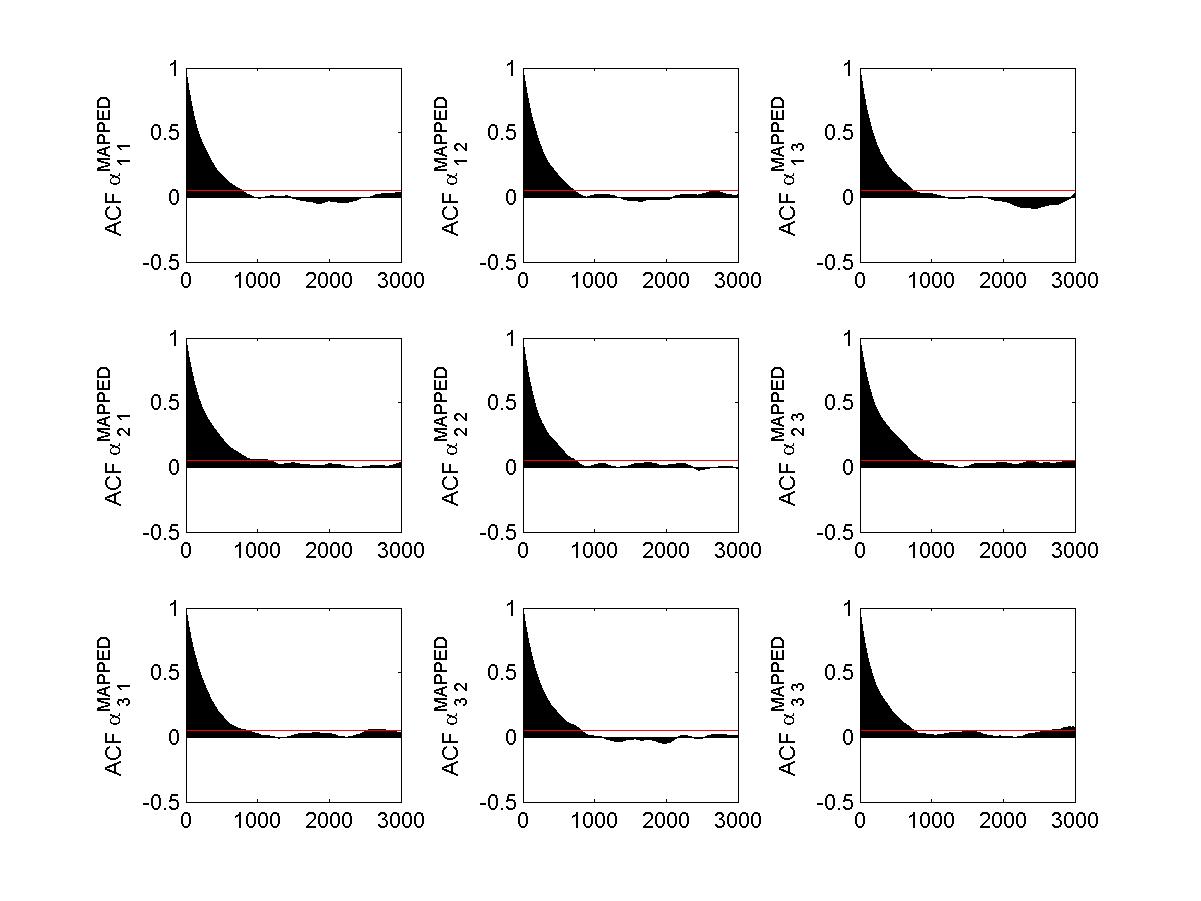}
\end{figure}

\begin{figure}[H]
\caption{\label{fig:4.2.8} The upper left plot shows the traces of total number
of components and of the number of active components at each step.
The lower left plot shows the corresponding running averages. The
plot on the right shows the traces of the mixture weights.}

\centering{}\includegraphics[scale=0.15]{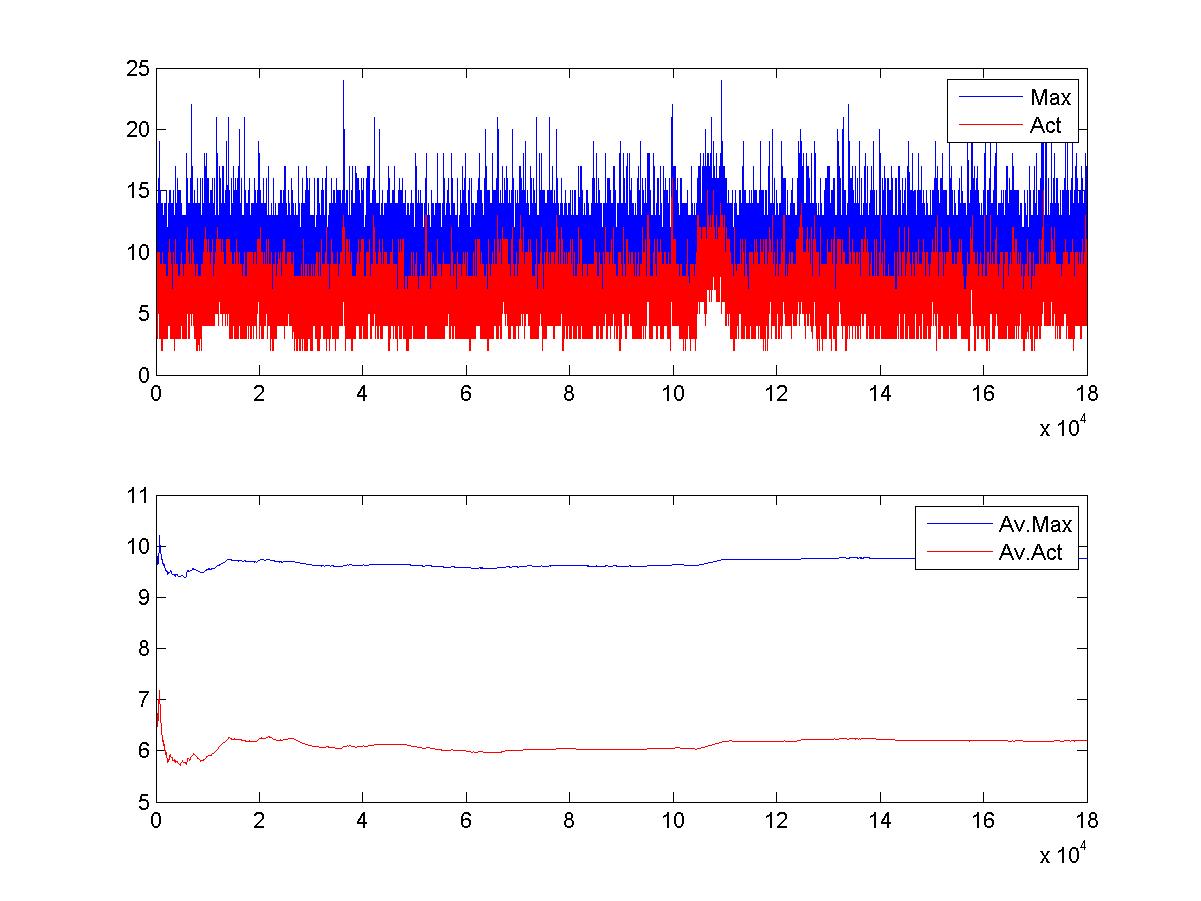}\includegraphics[scale=0.15]{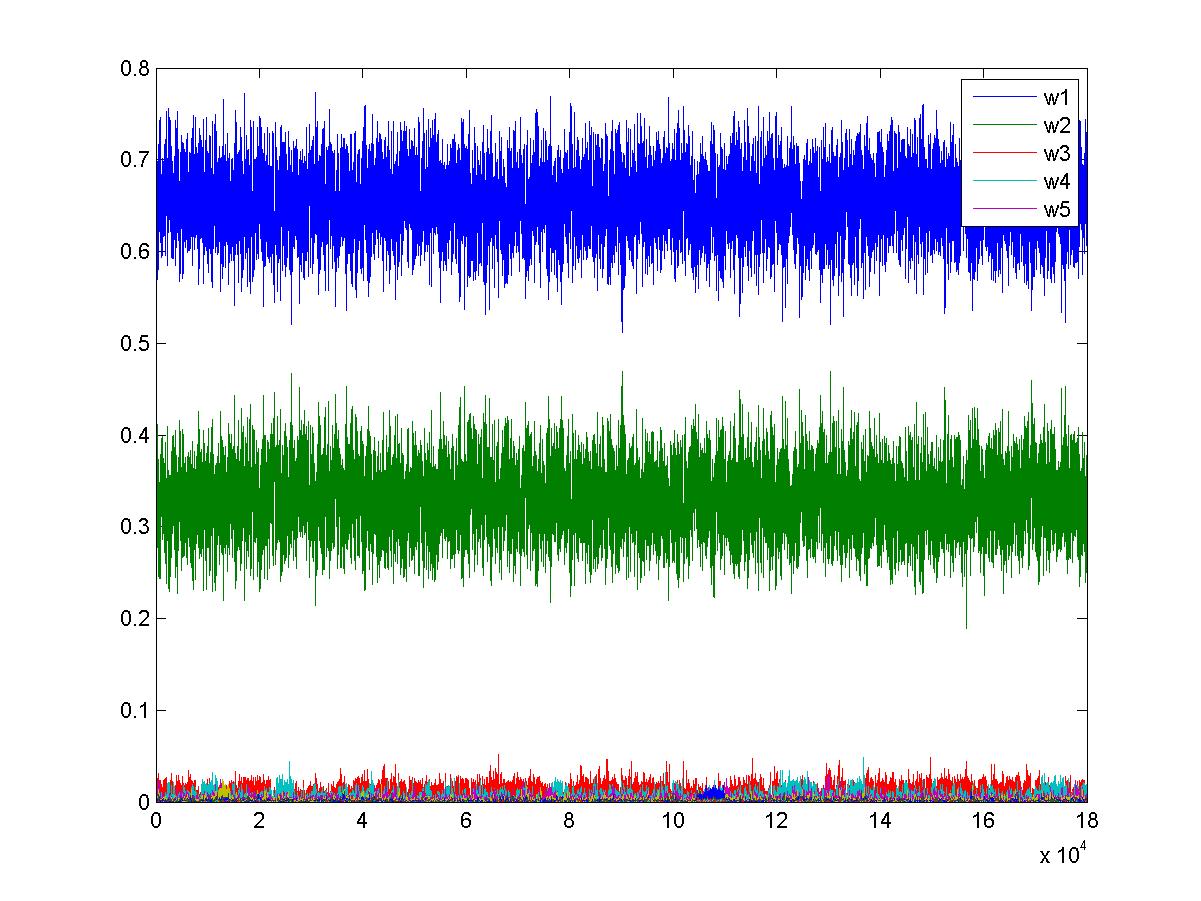}
\end{figure}

\begin{figure}[h]
\caption{\label{fig:4.2.9}True and estimated marginal densities of the innovations.}

\includegraphics[scale=0.12]{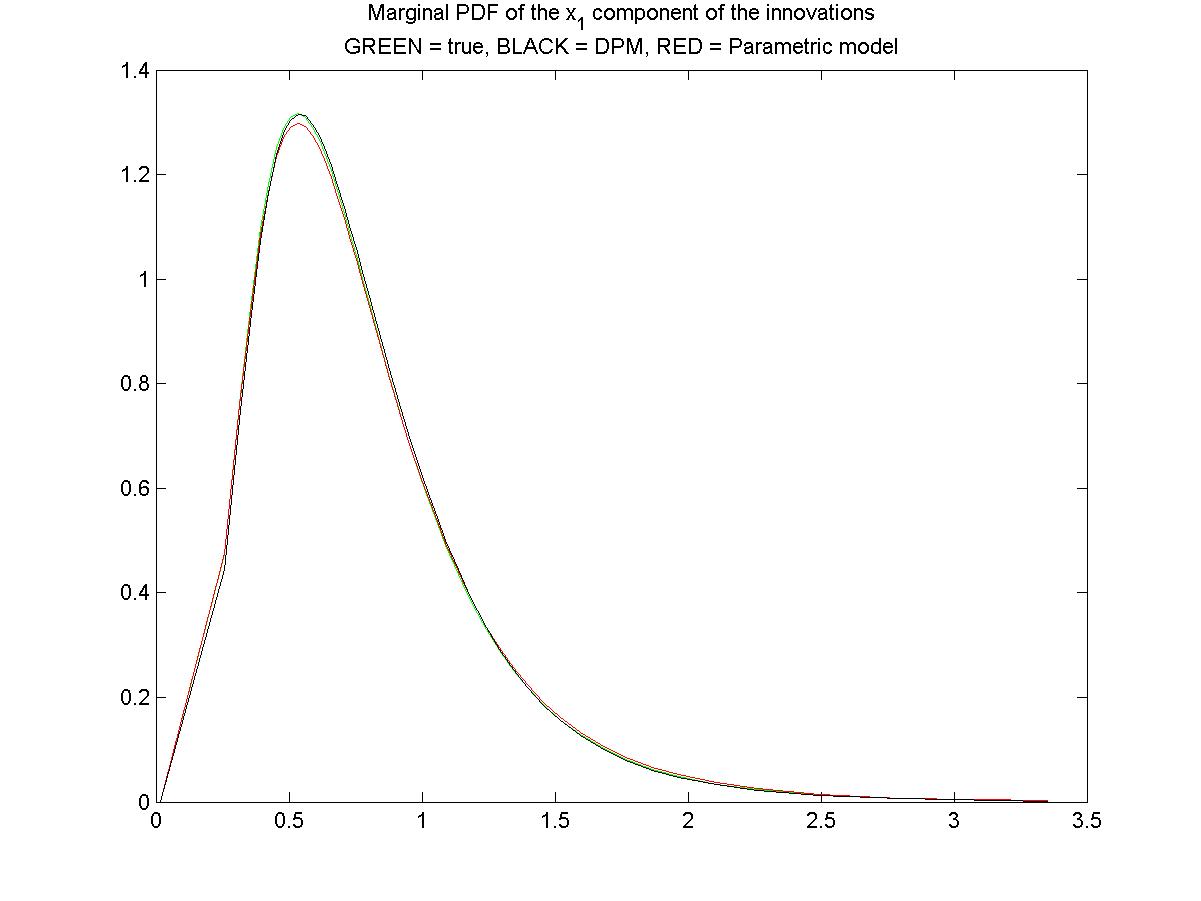}\includegraphics[scale=0.12]{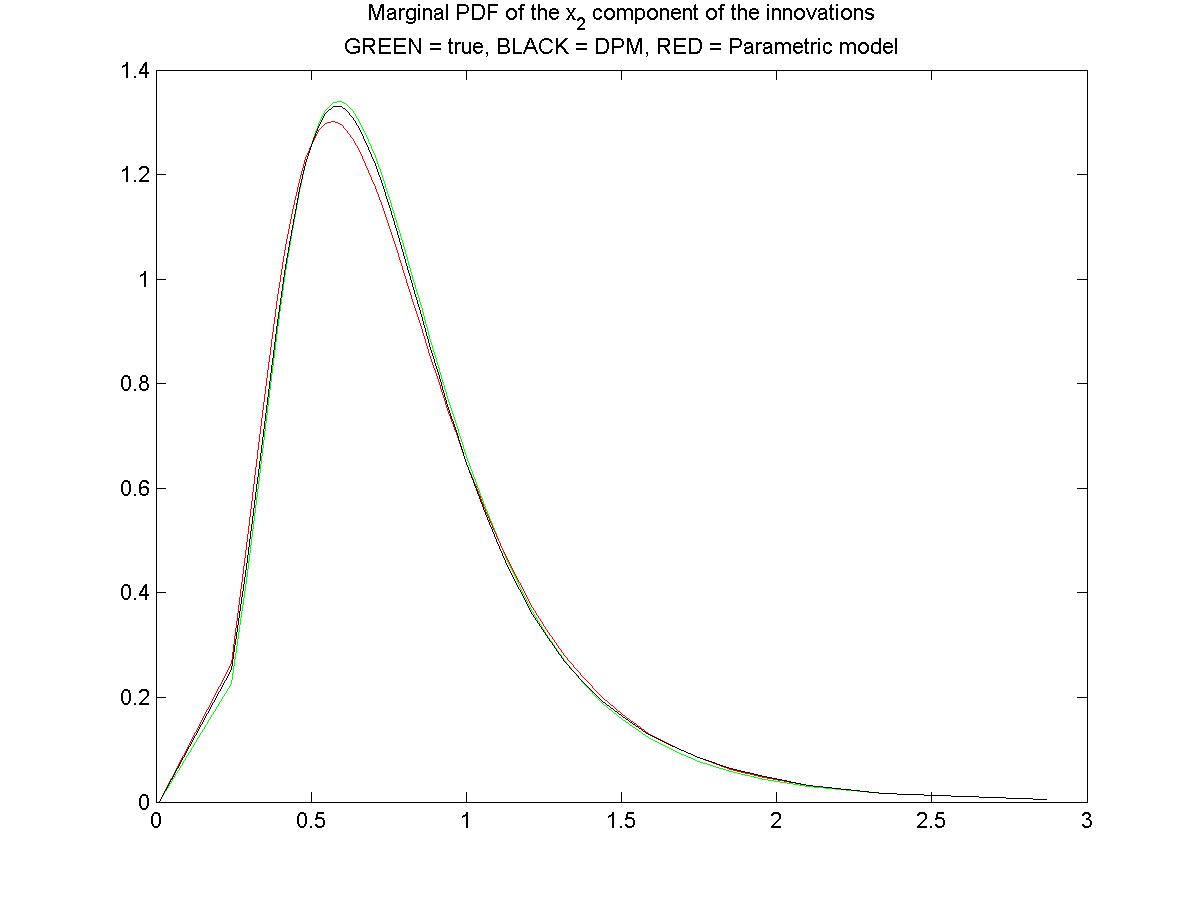}\includegraphics[scale=0.12]{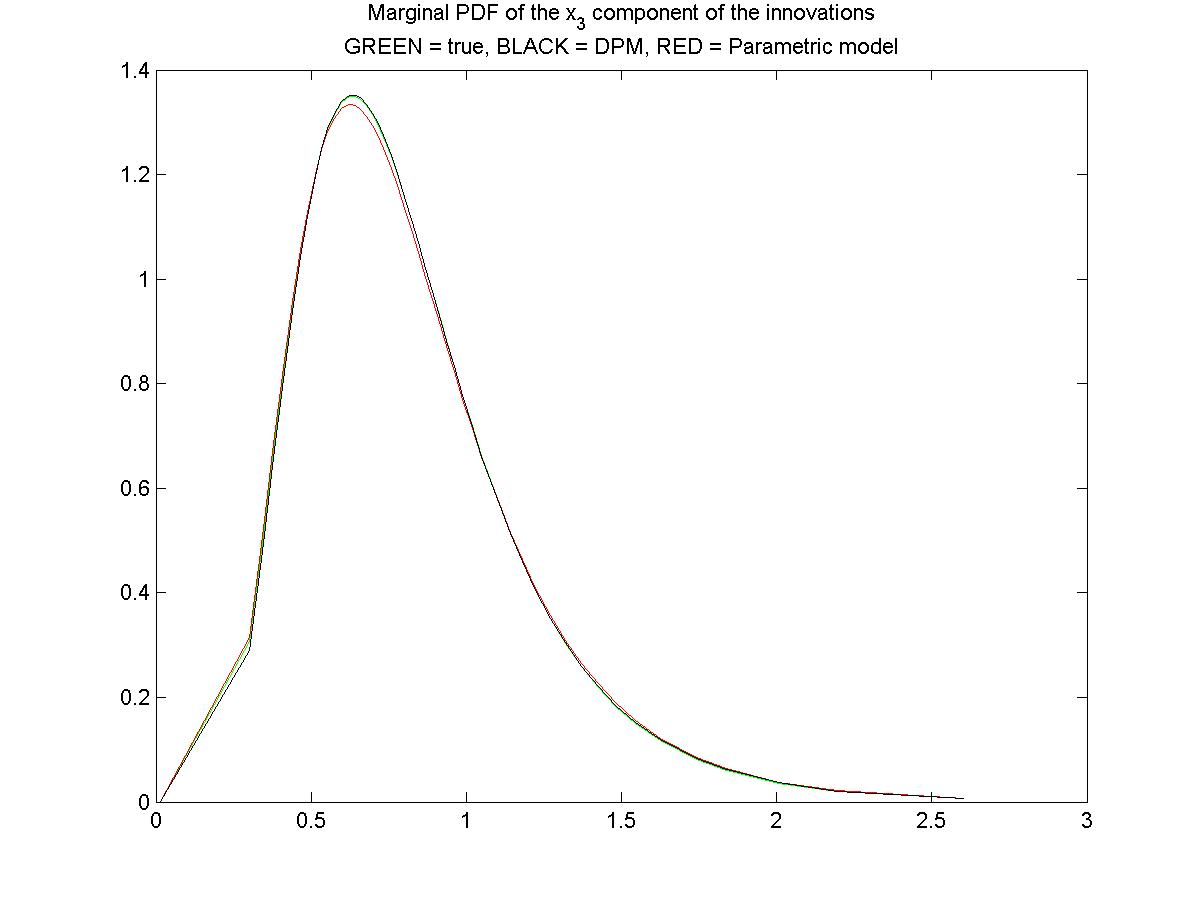}
\end{figure}

The concentration parameter of the DP
is $\alpha=1$, the truncation level defined in equation \eqref{eq:3.3.5}
is $\varepsilon_{\bar{\mathbf{m}}}=10^{-6}$ (different tolerance levels up to $10^{-10}$ returned no significant difference in the MCMC outputs), the initial value of
the vector of parameters of the conditional mean, $\bm{\eta}_{0}$,
is the maximum likelihood estimate found assuming a parametric model
with log-Normal distributed innovations and the parameters of the
base Normal-Wishart measure defined in equation \eqref{eq:3.4.4}
are 
\[
a=10+d,\ \mathbf{W}_{ij}=\begin{cases}
1 & if\ i=j\\
0 & else
\end{cases},\ \bm{\nu}=\bm{0},\ n_{0}=1
\]
where $\bm{\epsilon}_{t}=\mathbf{x}_{t}\oslash\bm{\mu}_{t}\left(\bm{\eta}_{0}\right)$,
$d=3$ is the length of vector $\mathbf{x}_{t}$ and the over-line
indicates sample mean over $t$. We sample 3000 trivariate observations from
the base vMEM specification with
\begin{align*}
\bm{\omega}=\left[\begin{array}{c}
0.35\\
0.59\\
0.43
\end{array}\right],\ \mathbf{B}=\left[\begin{array}{ccc}
0.36 & 0.07 & 0.18\\
0.20 & 0.24 & 0.14\\
0.01 & 0.10 & 0.41
\end{array}\right],\ \mathbf{A}=\left[\begin{array}{ccc}
0.21 & 0.14 & 0.04\\
0.13 & 0.28 & 0.09\\
0.07 & 0.08 & 0.30
\end{array}\right]
\end{align*}
 and 
 \small
\[
\bm{\varepsilon}_{t}\overset{{\scriptstyle i.i.d.}}{\sim}0.7*\text{logN}_{2}\left(\left[\begin{array}{c}
-0.200\\
-0.175\\
-0.150
\end{array}\right],\left[\begin{array}{ccc}
0.40 & 0.30 & 0.20\\
0.30 & 0.35 & 0.25\\
0.20 & 0.25 & 0.30
\end{array}\right]\right)+0.3*\text{logN}_{2}\left(\left[\begin{array}{c}
-0.185\\
-0.195\\
-0.125
\end{array}\right],\left[\begin{array}{ccc}
0.37 & 0.15 & 0.24\\
0.15 & 0.39 & 0.18\\
0.24 & 0.18 & 0.25
\end{array}\right]\right).
\]
\normalsize
We run the algorithm for $N_{it}=200,000$ iterations and we discard the
first 20,000 as burn-in. To sample the parameters of the conditional
mean we set $p=0.9$ as the weight for the proposal mixture density
and $\sigma_{1}=1$, $\sigma_{2}=\sqrt{21}$ as its scale factors.
The simulation time on a server running at 2.60GHz and with 128GB
RAM is about 13 hours. We finally repeat the whole procedure for 40 datasets. In Table \ref{tab:4.6} we report the posterior
means for the parameters of the conditional
mean, along with their true values, averaged over the 40 datasets, and with related 95\% credible intervals. As it can be seen, again, all
the true values of the parameters lie inside the 95\% 
intervals. All the estimates are based on effective sample sizes greater
than 327. In Figures \ref{fig:4.2.5}, \ref{fig:4.2.6}, \ref{fig:4.2.7}
we reported the traces, the posterior densities and the autocorrelation
functions of the post-processed parameters of the conditional mean for a randomly chosen dataset.
These figures show that all the traces have reached convergence and
all the autocorrelation functions become non-significant in less than
3000 lags, and most of them in less than 2000. In Figure \ref{fig:4.2.8}
are reported the traces and the running averages of the maximum number
of components and of the number of active components, at each step,
along with the traces of the mixture weights. We can see that there
are on average 6 active components but, correctly, only two of them
have really significant weights. Finally in Figure \ref{fig:4.2.9}
we report the true marginal densities of the innovations together
with the estimated marginal densities obtained with DMPLN2-vMEM and
the LN1-vMEM. Also in this case, the approximation obtained with the
DPMLN2-vMEM is better than the one obtained with the LN1-vMEM.

The repetitions over the 40 different datasets, whose figures are not reported for brevity, provide the same results: BSP-vMEM performs better than the simple parametric model in approximating the pdf of the innovations and the OSA pdf, and same convergence behaviour for the model parameters. Same considerations hold when we compare the proposed method with the Maximum a Posteriori of the Bayesian model with no DPM on the innovation error: the Log Pseudo Marginal Likelihood, estimated as suggested in \citet{nieto2014bayesian}, and averaged over the datasets, is equal to $-9.0578$ for our model versus $-9.1023$, therefore confirming our better performance. 

\begin{table}[h]
\caption{\label{tab:4.6}Posterior mean and 95\% quantiles for the
parameters of the conditional mean, averaged over 40 datasets.}

\centering{}%
\begin{tabular}{|c|c|c|c||c|c|c|c|}
\hline 
 & True & Est. & {\scriptsize{}(95\% C.I.)} &  & True & Est. & {\scriptsize{}(95\% C.I.)}\tabularnewline
\hline 
\hline 
$\omega_{1}$ & 0.35 & 0.3181 & {\scriptsize{}$\left(0.1267,0.6114\right)$} & $\alpha_{11}$ & 0.21 & 0.2087 & {\scriptsize{}$\left(0.1618,0.2586\right)$}\tabularnewline
\hline 
$\omega_{2}$ & 0.59 & 0.5664 & {\scriptsize{}$\left(0.4057,0.7725\right)$} & $\alpha_{21}$ & 0.13 & 0.1304 & {\scriptsize{}$\left(0.0865,0.1738\right)$}\tabularnewline
\hline 
$\omega_{3}$ & 0.43 & 0.3985 & {\scriptsize{}$\left(0.2908,0.5542\right)$} & $\alpha_{31}$ & 0.07 & 0.0693 & {\scriptsize{}$\left(0.0243,0.1013\right)$}\tabularnewline
\hline 
$\beta_{11}$ & 0.36 & 0.2796 & {\scriptsize{}$\left(0.0347,0.5526\right)$} & $\alpha_{12}$ & 0.14 & 0.1376 & {\scriptsize{}$\left(0.0763,0.1964\right)$}\tabularnewline
\hline 
$\beta_{21}$ & 0.10 & 0.0550 & {\scriptsize{}$\left(-0.2238,0.3659\right)$} & $\alpha_{22}$ & 0.28 & 0.2671 & {\scriptsize{}$\left(0.2259,0.3088\right)$}\tabularnewline
\hline 
$\beta_{31}$ & 0.01 & -0.0387 & {\scriptsize{}$\left(-0.3587,0.1651\right)$} & $\alpha_{32}$ & 0.08 & 0.0805 & {\scriptsize{}$\left(0.0389,0.1292\right)$}\tabularnewline
\hline 
$\beta_{12}$ & 0.07 & 0.1227 & {\scriptsize{}$\left(-0.1564,0.3755\right)$} & $\alpha_{13}$ & 0.04 & 0.0446 & {\scriptsize{}$\left(-0.0136,0.0957\right)$}\tabularnewline
\hline 
$\beta_{22}$ & 0.24 & 0.2963 & {\scriptsize{}$\left(-0.0396,0.5919\right)$} & $\alpha_{23}$ & 0.09 & 0.0947 & {\scriptsize{}$\left(0.0231,0.1610\right)$}\tabularnewline
\hline 
$\beta_{32}$ & 0.10 & 0.1705 & {\scriptsize{}$\left(-0.0508,0.4429\right)$} & $\alpha_{33}$ &  0.30 & 0.2960 & {\scriptsize{}$\left(0.2471,0.3524\right)$}\tabularnewline
\hline 
$\beta_{13}$ & 0.18 & 0.2039 & {\scriptsize{}$\left(0.0297,0.4187\right)$} & \multicolumn{1}{c}{} & \multicolumn{1}{c}{} & \multicolumn{1}{c}{} & \multicolumn{1}{c}{}\tabularnewline
\cline{1-4} 
$\beta_{23}$ & 0.14 & 0.1357 & {\scriptsize{}$\left(-0.0348,0.4135\right)$} & \multicolumn{1}{c}{} & \multicolumn{1}{c}{} & \multicolumn{1}{c}{} & \multicolumn{1}{c}{}\tabularnewline
\cline{1-4} 
$\beta_{33}$ & 0.41 & 0.3923 & {\scriptsize{}$\left(0.1534,0.5767\right)$} & \multicolumn{1}{c}{} & \multicolumn{1}{c}{} & \multicolumn{1}{c}{} & \multicolumn{1}{c}{}\tabularnewline
\cline{1-4} 
\end{tabular}
\end{table}

\section{Empirical analyses}

Volatility measurements using intra-daily data were first adopted by \citet{Parkinson1980Extreme} for the estimation of the daily range. Since then the literature has significantly expanded: from the realized volatility of \citet{Andersen1998Answering} and \citet{peluso2019conditionally}  to realized
kernels of \citet{Barndorff2008Designing} and realized covariance matrices \citep{ait2010high,peluso2014bayesian,corsi2015missing}. In parallel to
the evolution of these measures, there has been a natural complementary
effort to build adequate models to describe their dynamics. Multiplicative
Error Models have been used for this purpose for example by \citet{Cipollini2013SemiparametricVMEM}.
The time series which are most commonly used in this respect are the
squared close-to-close adjusted returns $r_{t}^{2}$, the realized
variances $rv_{t}^{2}$ (in any of their flavours), the absolute returns
$\left|r_{t}\right|$, the realized volatilities $rv_{t}$, and the
daily ranges $hl_{t}$. We will now illustrate the characteristics
of our DPMLN2-vMEM with 
\[
\bm{\mu}_{t}=\bm{\omega}+\bm{\beta}\bm{\mu}_{t-1}+\mathbf{A}\mathbf{x}_{t-1}
\]
in modelling the interaction among several volatility measures for
the purpose of forecasting, with DP concentration parameter $\alpha=1$ as in the simulation study. We will make a comparison between the
DPMLN2-vMEM and the LN1-vMEM, estimated using MAP (Maximum A Posteriori), in terms of their
(in the sample) predictive performance. To do this we will use the
Log-Predictive Score (LPS) proposed by \citet{Kim1998Stochastic},
defined as:
\[
LPS=-\frac{1}{T}\underset{{\scriptstyle t=1}}{\overset{{\scriptstyle T}}{\sum}}\log\hat{f}_{\mathbf{x}_{t}}\left(\mathbf{x}_{t}\right)=-\frac{1}{T}\underset{{\scriptstyle t=1}}{\overset{{\scriptstyle T}}{\sum}}\log\left(\underset{{\scriptstyle h=1}}{\overset{{\scriptstyle d}}{\prod}}\frac{1}{\hat{\mu}_{h}^{\left(t\right)}}\hat{f}_{\bm{\varepsilon}_{t}}\left(\mathbf{x}_{t}\oslash\hat{\bm{\mu}}_{t}\right)\right)
\]
where the probability density function of the innovations has been estimated as:
\begin{equation}
\hat{f}_{\bm{\varepsilon}_{t}}\left(e\right)=\frac{1}{N_{it}}\underset{{\scriptstyle n=1}}{\overset{{\scriptstyle N_{it}}}{\sum}}\underset{{\scriptstyle j=1}}{\overset{{\scriptstyle N^{\left(n\right)}}}{\sum}}w_{j}^{\left(n\right)}\text{logN}_{d}\left(e;\mathbf{m}_{j}^{\left(n\right)},\bm{\Sigma}_{j}^{\left(n\right)}\right)
\label{eq:5.1}
\end{equation}
and $N^{\left(n\right)}$ is the number of components that appear in the n-th
MCMC step. It is such that $\underset{{\scriptstyle j=1}}{\overset{{\scriptstyle N^{\left(n\right)}}}{\sum}}w_{j}^{\left(n\right)}>0.99$ for all $n=1,\ldots,N_{it}$. Thus a lower LPS is an indication of a better predictive
performance. We further compare the methods according to a second measure of performance: the Log Pseudo Marginal Likelihood (LPML), estimated as in \citet{nieto2014bayesian}, and without the negative sign, to be coherent in interpretation to LPS that a lower value corresponds to a better performance.

\begin{figure}
\caption{\label{fig:5.1.1}Time series of realized volatilities and absolute
returns of S\&P 500, DJIA, FTSE 100}

\centering{}\includegraphics[scale=0.2]{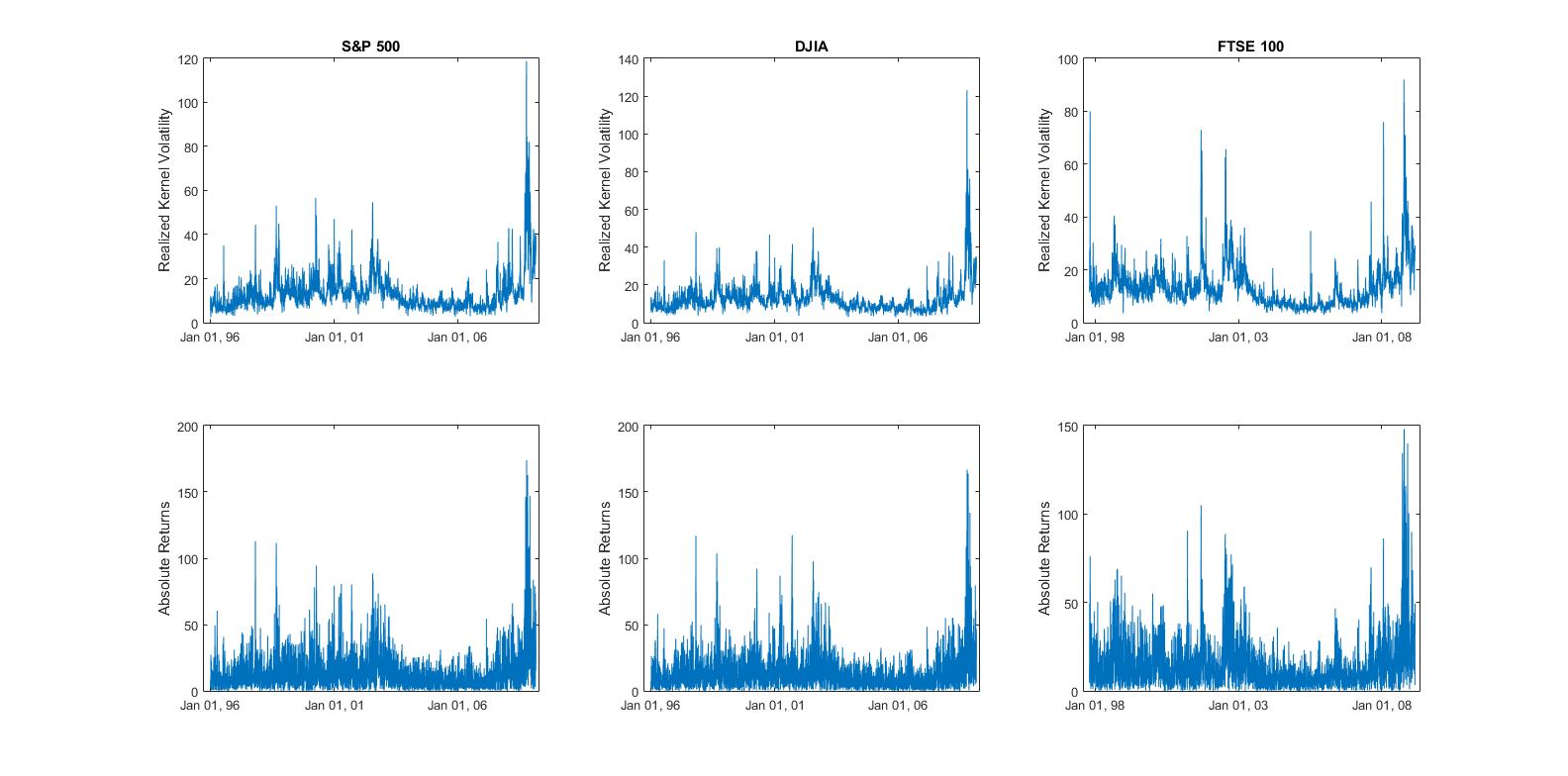}
\end{figure}

For our analysis we make use of a bivariate series composed
by daily absolute returns and realized kernel volatilities, $\left(\left|r_{t}\right|,rv_{t}\right)$.
We take the data from the Oxford Man Institute ``Realized Library''
(\citet{Shephard2010Realising}) and we express them in annualized
percentage terms through the transformation:
\[
x_{t}^{AP}=100\sqrt{252}x_{t}.
\]
We run our analysis on three stock indices: Standard \& Poor 500 (S\&P
500), Dow Jones Industrial Average (DJIA), Financial Times Stock Exchange
100 (FTSE 100). The covered period is the one between January 1996
and February 2009 for a total of 3261 observations for the S\&P 500
series, 3260 observations for the DJIA series and 2840 observations
of the FTSE 100 series. From the time series plotted in Figure \ref{fig:5.1.1},
we can see that both the measures of all the three indices share some
common features like alternance of periods of high and low volatility
and persistence.

\begin{figure}[H]
\caption{\label{fig:5.2.1}MCMC traces, posterior histograms and ACFs of the
components of the post-processed vector $\bm{\omega}$. The green
lines in the histogram represent the 95\% C.I.}

\includegraphics[scale=0.12]{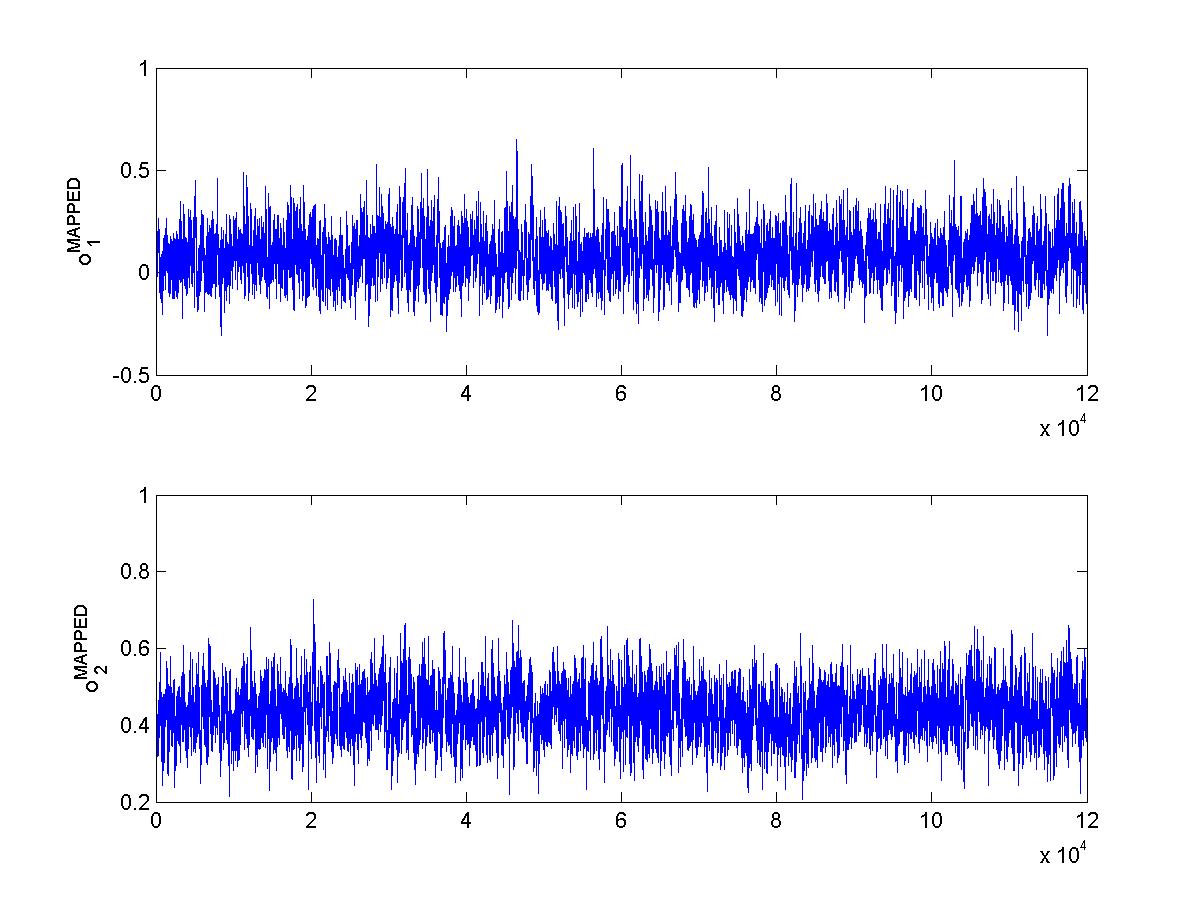}\includegraphics[scale=0.12]{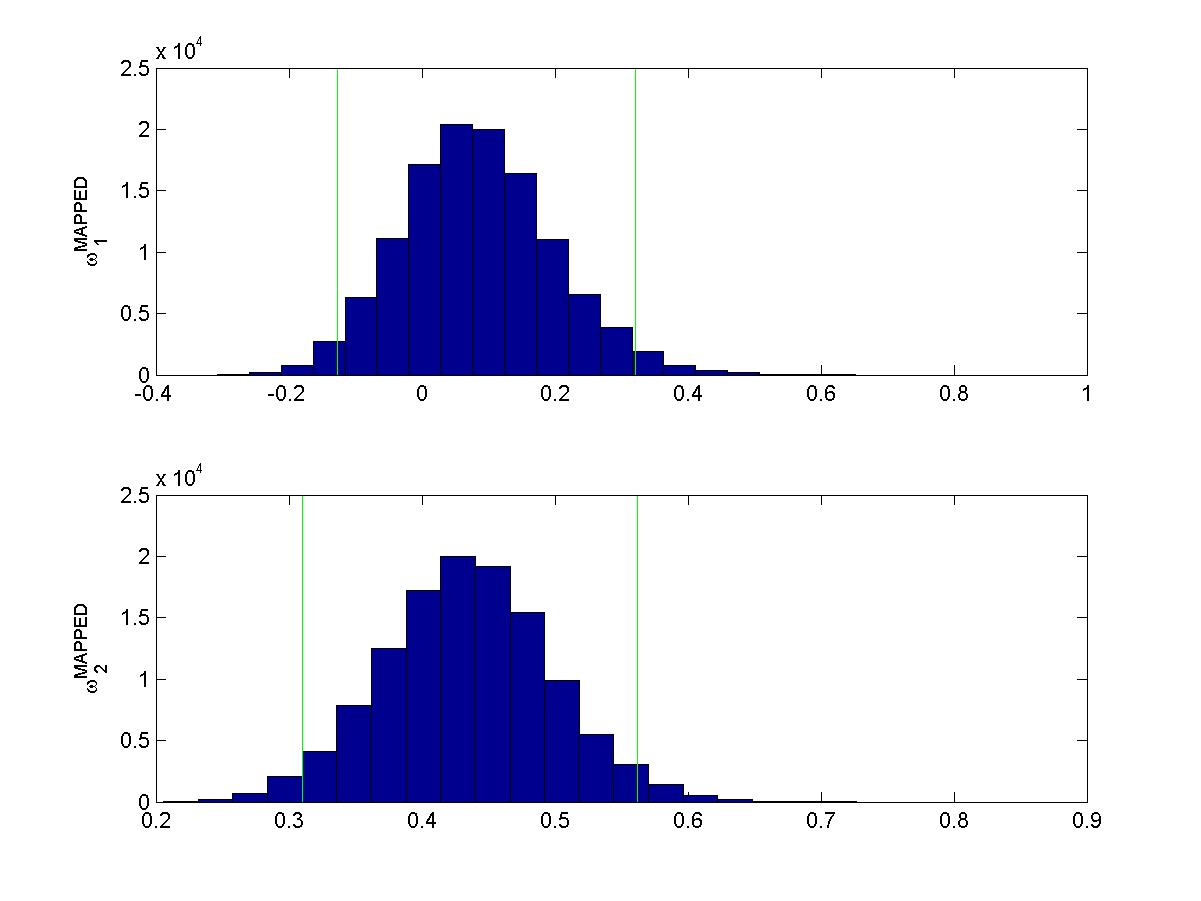}\includegraphics[scale=0.12]{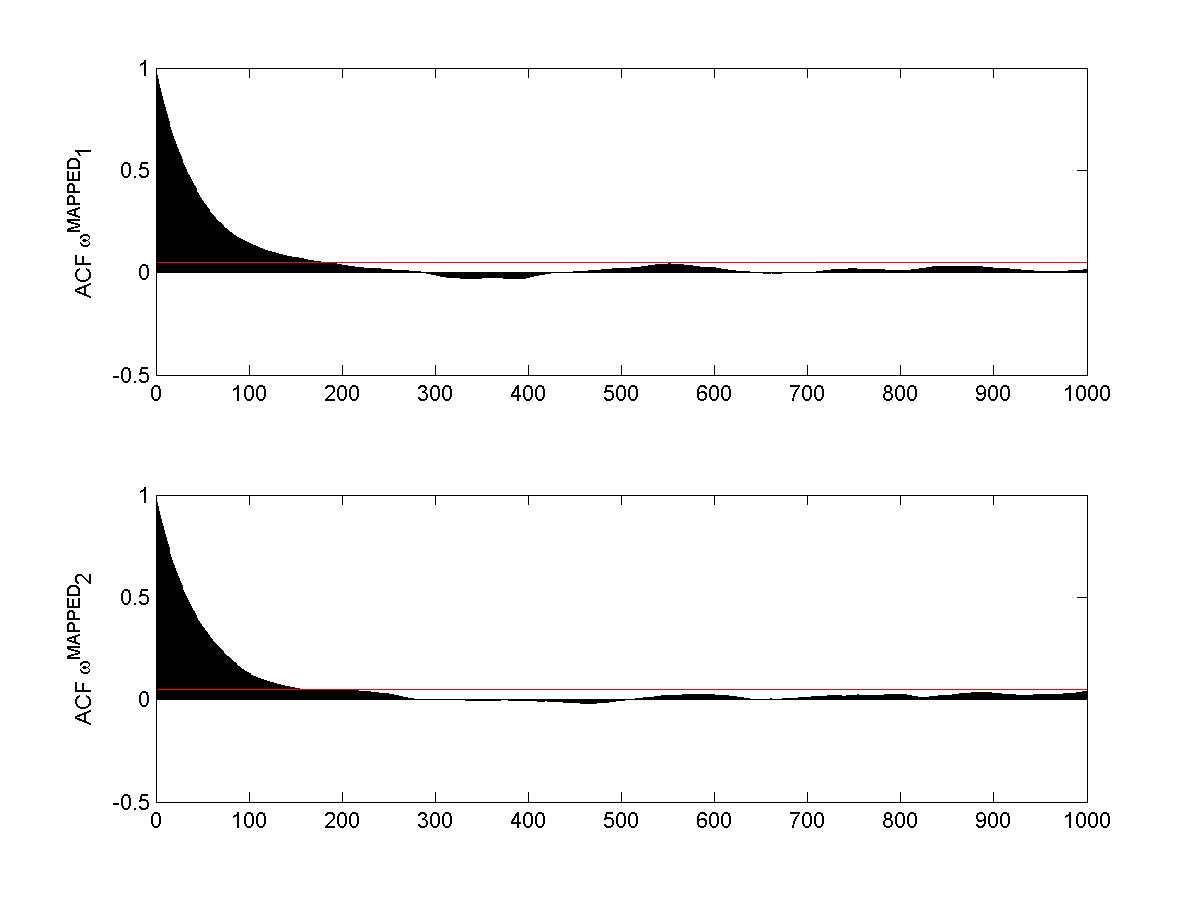}
\end{figure}

\begin{figure}[H]
\caption{\label{fig:5.2.2}Estimated joint and marginal densities of the innovations
over the estimated innovations. On the upper row there are the results
obtained with the DPMLN2-vMEM, on the lower row the ones obtained
with LN1-vMEM }

\begin{centering}
\includegraphics[scale=0.12]{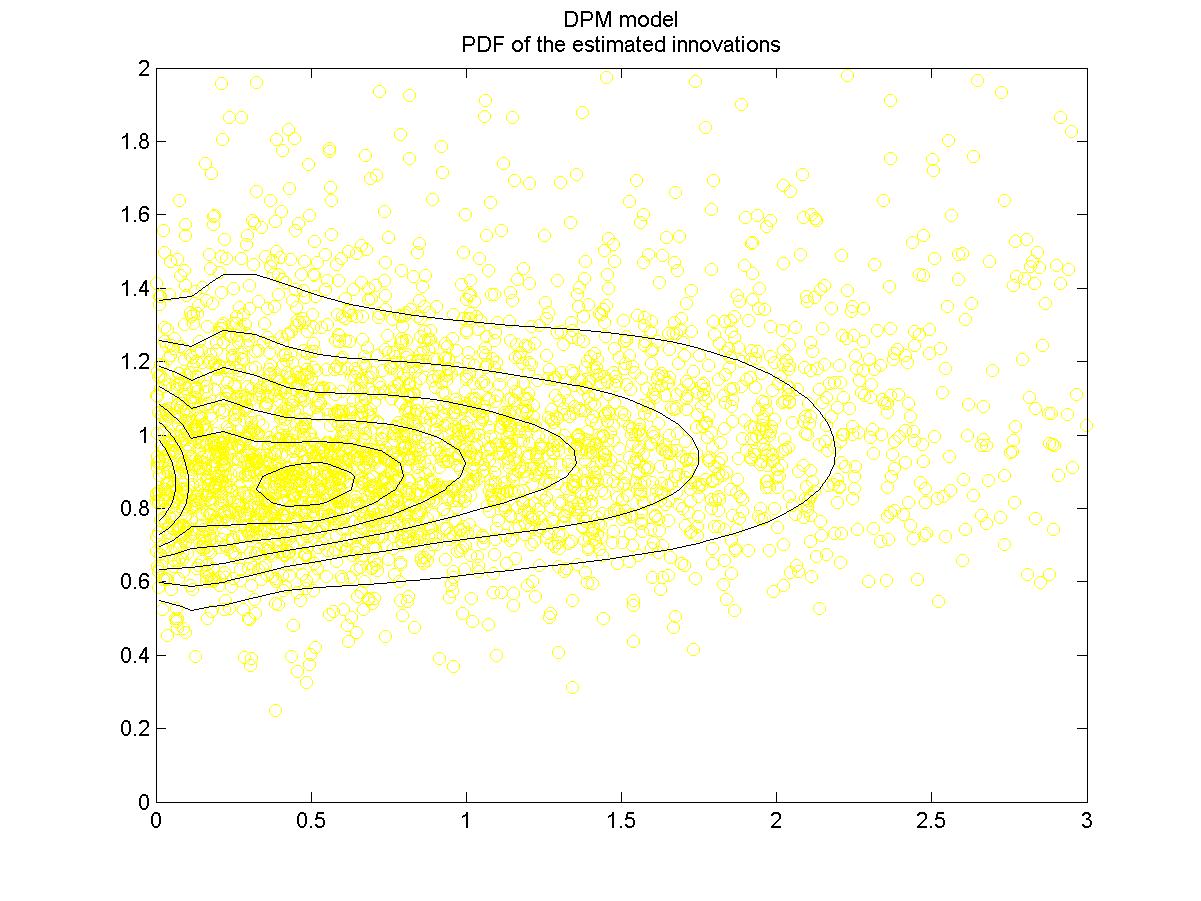}\includegraphics[scale=0.12]{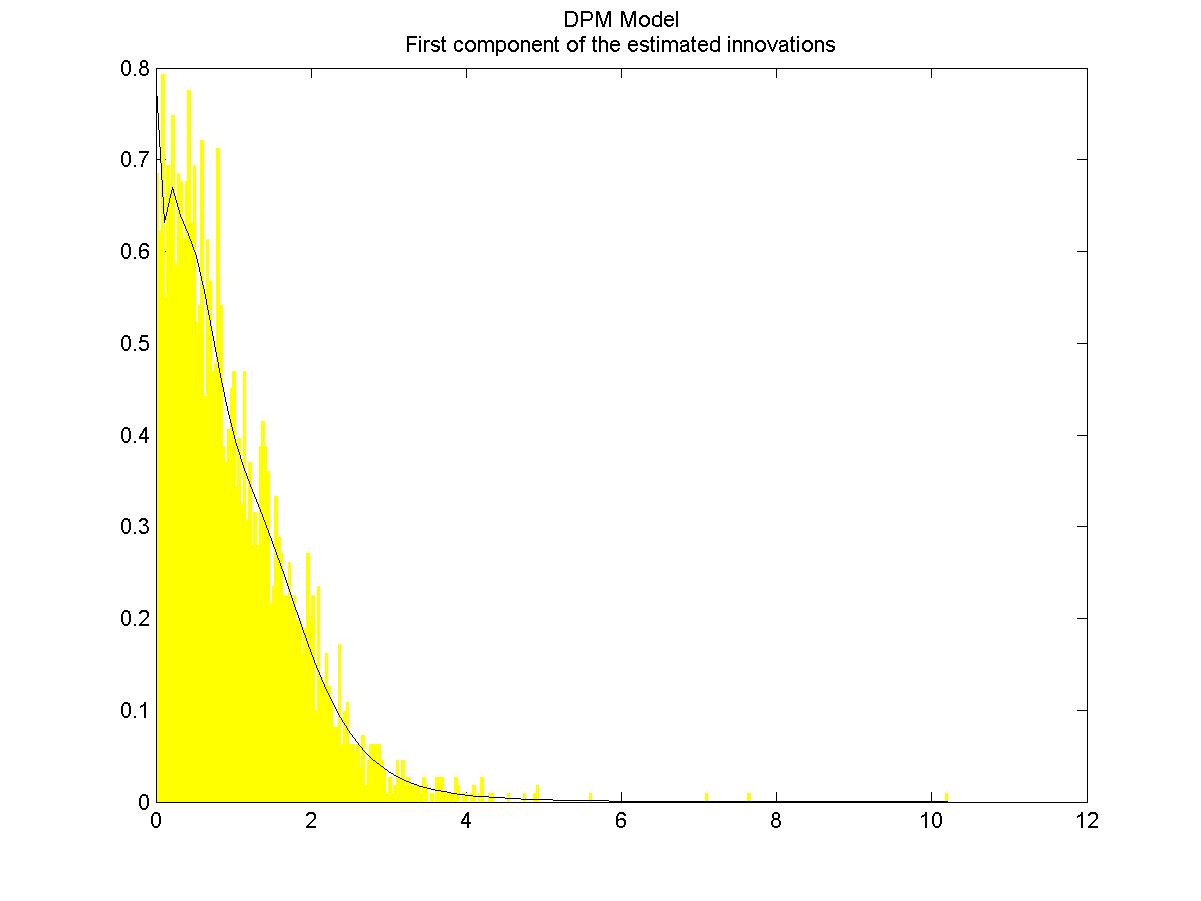}\includegraphics[scale=0.12]{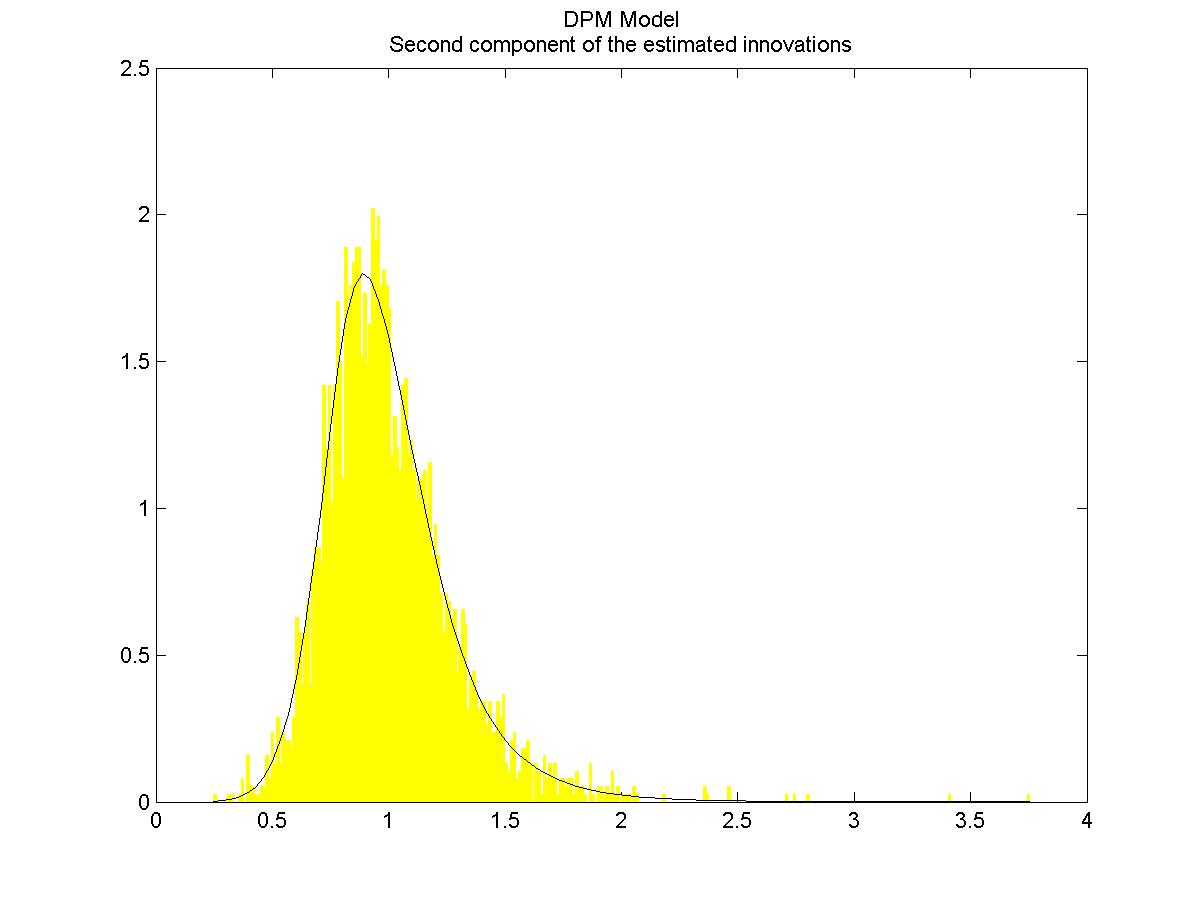}
\par\end{centering}

\centering{}\includegraphics[scale=0.12]{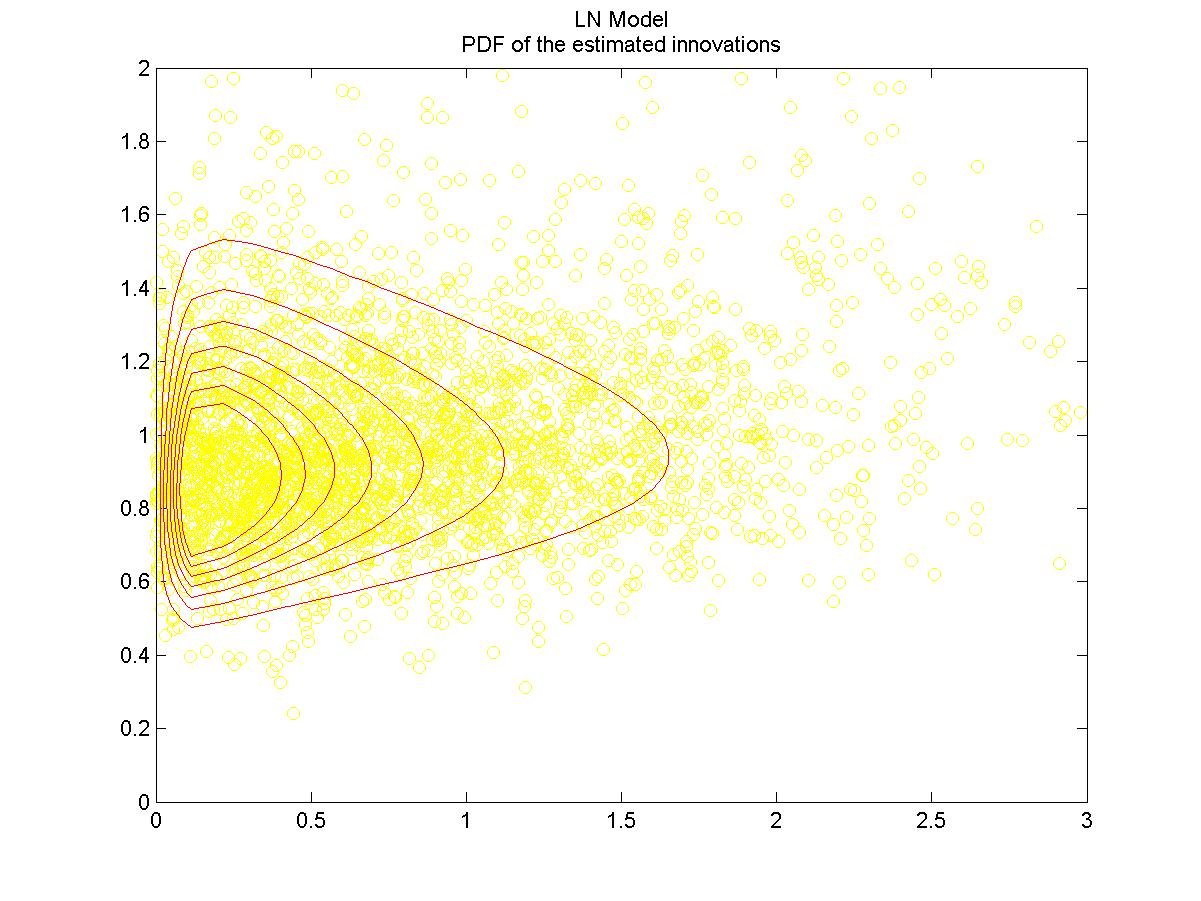}\includegraphics[scale=0.12]{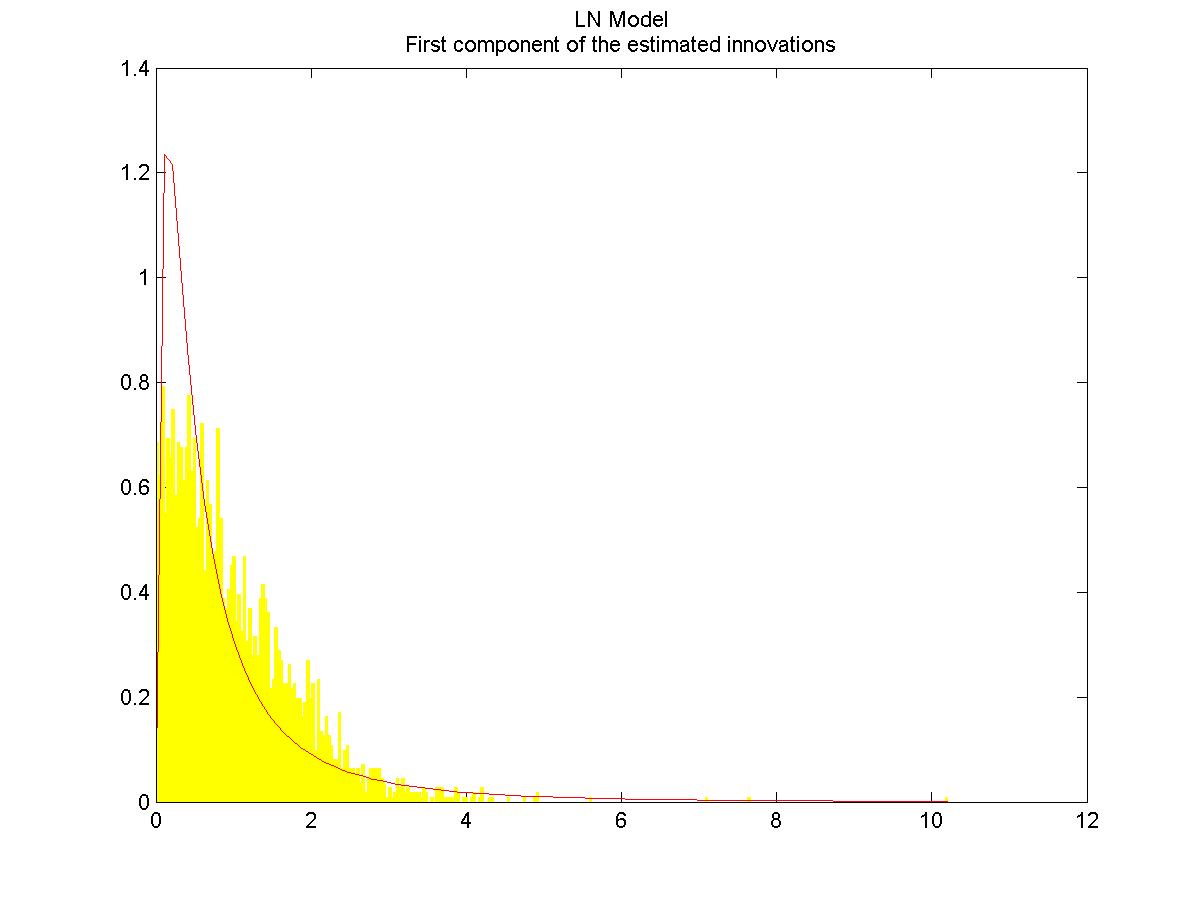}\includegraphics[scale=0.12]{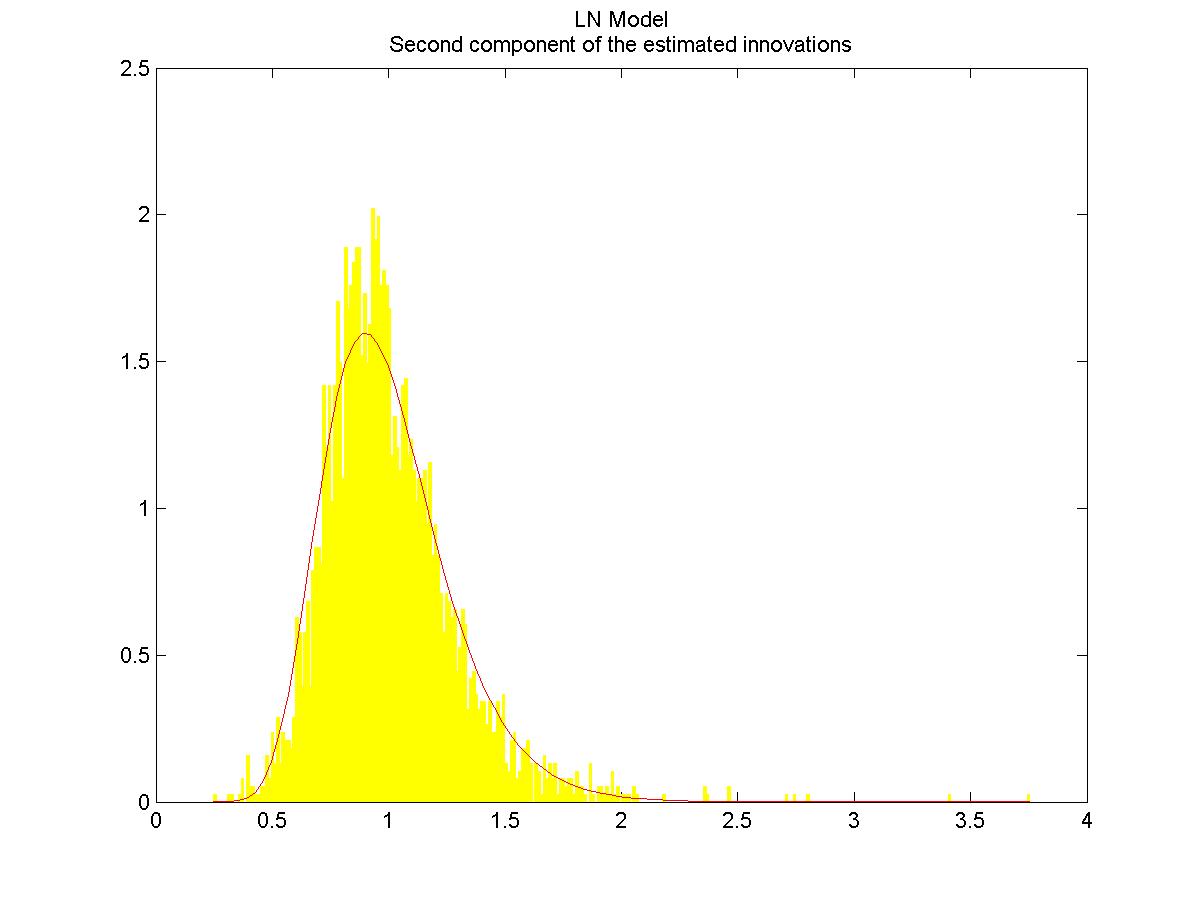}
\end{figure}

\begin{figure}[H]
\caption{\label{fig:5.2.2-1}The upper plot shows the traces of total number
of components and of the number of active components at each step.
The lower plot shows the corresponding running averages. }

\centering{}\includegraphics[scale=0.15]{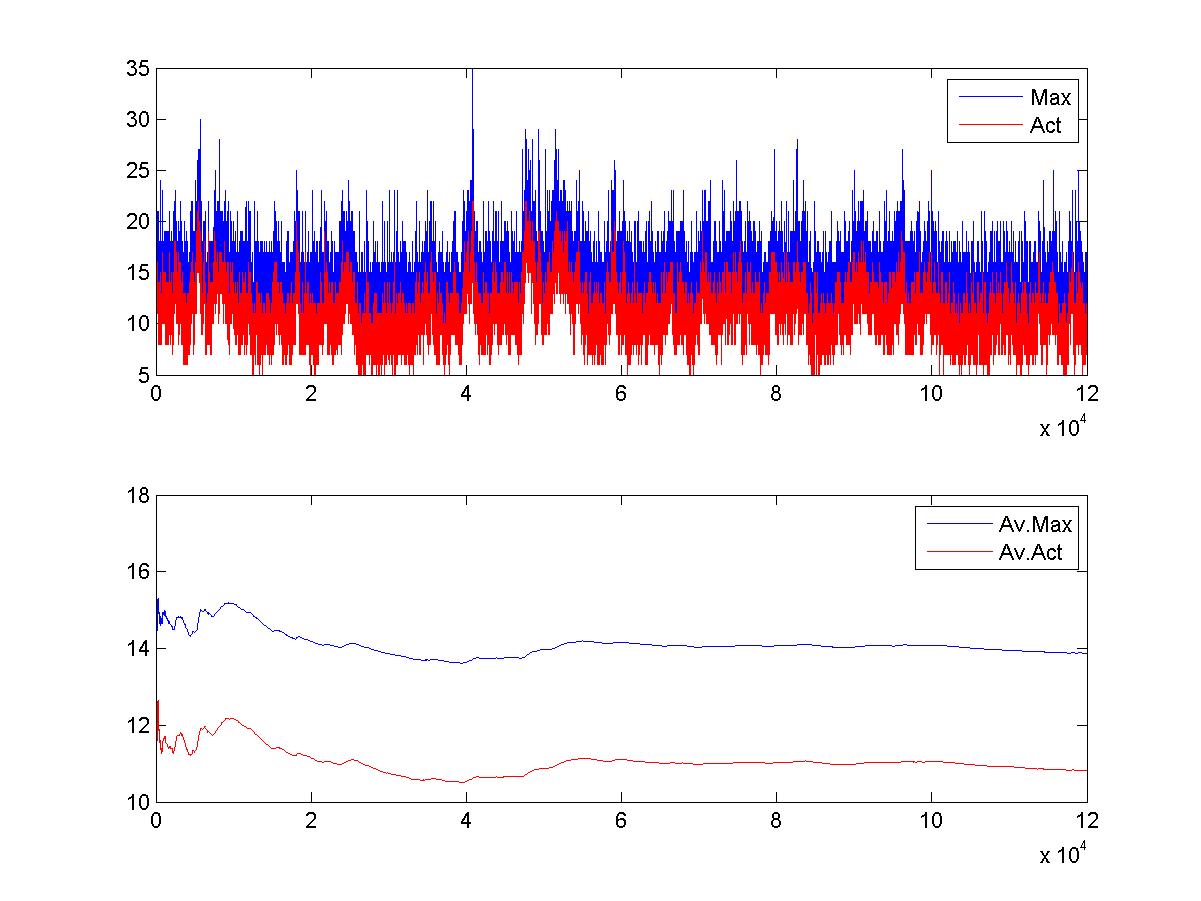}
\end{figure}

For all the time series, we run 150,000 iterations of the algorithm
described in Section 3 and then discard the first 30,000 of them as
burn-in. In Table \ref{tab:5.1} we report, for all the time series,
the estimates of the conditional mean parameters and the 95\% credible
intervals obtained with our model together with the maximum likelihood
estimates of the same parameters and the corresponding standard errors
obtained from the LN1-vMEM. In all the analyses, all the effective
sample sizes of the variables obtained from the MCMC simulations are
bigger than 500. In Table \ref{tab:5.2} we report the Log-Predictive
scores and Log-Pseudo Marginal Likelihoods of the DPMLN2-vMEM and the LN1-vMEM. For the sake of brevity,
we will plot hereafter only the traces, histograms and autocorrelation
functions of the elements of the $\bm{\omega}$ vector (Figure \ref{fig:5.2.1}),
the estimations of the joint density and its marginals (Figure \ref{fig:5.2.2})
and the traces of the total number of components (Figure \ref{fig:5.2.2-1})
obtained analysing the time series of the S\&P 500 index with DPMLN2-vMEM
and LN1-vMEM. Very similar figures have been obtained for other
parameters and for the other two time series.

\begin{table}[H]
\caption{\label{tab:5.1}Posterior mean, 95\% credible intervals, MAP estimates
and corresponding standard error, for the parameters of the conditional
mean.}

\begin{tabular}{|c||c|c||c|c||c|c|}
\multicolumn{1}{c}{} & \multicolumn{1}{c}{S\&P 500} & \multicolumn{1}{c}{} & \multicolumn{1}{c}{DJIA} & \multicolumn{1}{c}{} & \multicolumn{1}{c}{FTSE 100} & \multicolumn{1}{c}{}\tabularnewline
\hline 
 & MCMC Est. & MAP Est. & MCMC Est. & MAP Est. & MCMC Est. & MAP Est.\tabularnewline
 & {\scriptsize{}(95\% C.I.)} & {\scriptsize{}(S.d.)} & {\scriptsize{}(95\% C.I.)} & {\scriptsize{}(S.d.)} & {\scriptsize{}(95\% C.I.)} & {\scriptsize{}(S.d.)}\tabularnewline
\hline 
\hline 
$\omega_{1}$ & 0.0797 & -0.0361 & 0.0200 & -0.1158 & 0.0688 & -0.0486\tabularnewline
 & {\footnotesize{}$\left(-0.1290,0.3129\right)$} & {\footnotesize{}0.2475} & {\footnotesize{}$\left(-0.3079,0.3068\right)$} & {\footnotesize{}0.2527} & {\footnotesize{}$\left(-0.1119,0.2623\right)$} & {\footnotesize{}0.2156}\tabularnewline
\hline 
$\omega_{2}$ & 0.3984 & 0.4686 & 0.3963 & 0.4520 & 0.1580 & 0.2089\tabularnewline
 & {\footnotesize{}$\left(0.2834,0.5229\right)$} & {\footnotesize{}0.0604} & {\footnotesize{}$\left(0.2825,0.5171\right)$} & {\footnotesize{}0.0572} & {\footnotesize{}$\left(0.0914,0.2311\right)$} & {\footnotesize{}0.0401}\tabularnewline
\hline 
$\beta_{1}$ & 0.7217 & 0.6379 & 0.6220 & 0.6387 & 0.6940 & 0.6629\tabularnewline
 & {\footnotesize{}$\left(0.6505,0.7843\right)$} & {\footnotesize{}0.0524} & {\footnotesize{}$\left(0.5292,0.7055\right)$} & {\footnotesize{}0.0525} & {\footnotesize{}$\left(0.6223,0.7587\right)$} & {\footnotesize{}0.0624}\tabularnewline
\hline 
$\beta_{2}$ & 0.5797 & 0.5604 & 0.5722 & 0.5622 & 0.7078 & 0.6735\tabularnewline
 & {\footnotesize{}$\left(0.5404,0.6173\right)$} & {\footnotesize{}0.0157} & {\footnotesize{}$\left(0.5321,0.6113\right)$} & {\footnotesize{}0.0154} & {\footnotesize{}$\left(0.6700,0.7428\right)$} & {\footnotesize{}0.0125}\tabularnewline
\hline 
$\alpha_{11}$ & -0.0442 & -0.1131 & -0.0530 & -0.0925 & -0.0261 & -0.0574\tabularnewline
 & {\footnotesize{}$\left(-0.0686,-0.0202\right)$} & {\footnotesize{}0.0230} & {\footnotesize{}$\left(-0.0828,-0.0235\right)$} & {\footnotesize{}0.0251} & {\footnotesize{}$\left(-0.0556,0.0033\right)$} & {\scriptsize{}0.0282}\tabularnewline
\hline 
$\alpha_{21}$ & 0.0397 & 0.0394 & 0.0355 & 0.0369 & 0.0271 & 0.0326\tabularnewline
 & {\footnotesize{}$\left(0.0296,0.0502\right)$} & {\footnotesize{}0.0053} & {\footnotesize{}$\left(0.0262,0.0450\right)$} & {\footnotesize{}0.0048} & {\footnotesize{}$\left(0.0182,0.0361\right)$} & {\scriptsize{}0.0046}\tabularnewline
\hline 
$\alpha_{12}$ & 0.3353 & 0.5917 & 0.4518 & 0.5611 & 0.3533 & 0.5139\tabularnewline
 & {\footnotesize{}$\left(0.2651,0.4150\right)$} & {\footnotesize{}0.0754} & {\footnotesize{}$\left(0.3554,0.5564\right)$} & {\footnotesize{}0.0761} & {\footnotesize{}$\left(0.2694,0.4448\right)$} & {\footnotesize{}0.0970}\tabularnewline
\hline 
$\alpha_{22}$ & 0.3491 & 0.3625 & 0.3608 & 0.3641 & 0.2519 & 0.2758\tabularnewline
 & {\footnotesize{}$\left(0.3151,0.3845\right)$} & {\footnotesize{}0.0151} & {\footnotesize{}$\left(0.3252,0.3973\right)$} & {\footnotesize{}0.0145} & {\footnotesize{}$\left(0.2195,0.2870\right)$} & {\footnotesize{}0.0125}\tabularnewline
\hline 
\end{tabular}
\end{table}

\begin{table}[H]
\caption{\label{tab:5.2}Log-Predictive Scores (LPS) and Log Pseudo Marginal Likelihoods (LPML) for parametric and semiparametric models, for all the three series.}

\centering{}%
\begin{tabular}{|c|c|c|c|}
\hline 
 & S\&P 500 & DJIA & FTSE 100\tabularnewline
\hline 
\hline 
LN1-vMEM LPS & 6.1795 & 6.0238 & 6.1318\tabularnewline
\hline 
DPMLN2-vMEM LPS & 6.0176 & 5.8676 & 5.9124\tabularnewline
\hline 
LN1-vMEM LPML & 6.1882 & 6.0270 & 6.1349\tabularnewline
\hline 
DPMLN2-vMEM LPML & 6.0370 & 5.8952 & 5.9394\tabularnewline
\hline 
\end{tabular}
\end{table}

From Figure \ref{fig:5.2.1} it can be seen that, although
there is some autocorrelation the traces of the MCMC simulations have
all reached convergence and the posterior histograms look informative.

For all the three time series and both the models considered, we obtain
some common qualitative features of the point estimates. First of
all the $\beta$s are always the biggest coefficients, meaning that
the factor that influences the most the evolution of the conditional
mean is always its lagged realization. Second, the estimates of the
coefficients of the second column of matrix $\mathbf{A}$ are always
bigger, in absolute value, than the ones in the first column of the
same matrix. This suggests that the lagged realizations of the realized
kernel volatility influence the evolution of both the components of
the conditional mean vector more than the lagged realizations of the
absolute returns. This fact, that could look strange at first sight,
simply means that the lagged observation of the realized volatility
contains more information on the present realization of the conditional
mean of the absolute returns than the lagged absolute returns and
this can be viewed as a further proof of the fact that the realized
volatilities are more informative about the latent volatility than
the absolute returns. Furthermore the $\alpha_{11}$ coefficient is
always negative meaning that the conditional mean of the absolute
returns depends inversely from the lagged realizations of the absolute
returns.

In all the empirical analyses we tried there have always been about
the same average number of active and total components of the DPM,
for all the time series. Furthermore for all the time series there
are always at least eight active components of the DPM for the whole
MCMC run. The sum of the average of the weights of these components
is always bigger than 0.99, meaning that, even in the MCMC steps in
which there are more components, the first eight are dominant.

Regarding the approximation of the distribution of the data obtained
with the DPMLN2-vMEM and with the LN1-vMEM, Figure \ref{fig:5.2.2}
suggets that our semiparametric model outperforms its parametric counterpart.
This can be perceived from the graphs of the joint distribution, especially near the $y$-axis, but
it becomes clearer looking at the graphs of the marginals: whilst the second marginals are indeed very similar, the approximation of the first marginal (second column of the figure) obtained with
DPMLN2-vMEM is much better than the one obtained with LN1-vMEM, since the DPM model seems to better approximate the data close to zero. 
For what it takes the predictive performances in the sample, Table \ref{tab:5.2} suggests that DPMLN2-vMEM performs better than LN1-vMEM.

\section{Conclusions}

We proposed a Bayesian semiparametric vMEM for non-negative multivariate random vectors, a relevant setting in many financial applications. 
Our contribution is the formulation of a statistical model that is (i) not bounded to special parametric forms of the error term distribution, known to be a quite strong restriction with multivariate data, and (ii) subject to weaker assumptions than the other semiparametric approaches in the literature, based on the Generalized Method of Moments.

In more details, the innovation term of our vMEM follows a location-scale DPM of multivariate log-normal distributions. 
By exploiting a parameter-expanded unconstrained version of the model, we are able to simplify the computational difficulties arising from the constraints to the positive orthant and we formalize an efficient slice sampler for posterior inference. 
The proposed model shows better fitting and predictive performances than its parametric counterpart in both the simulations and in the empirical study on the interaction between daily absolute returns and realized kernel volatilities. 

Further developments of interest include 
(i) a refinement of the sampling technique with sparsity-driven efficiencies that can manage time series in high dimension, 
(ii) a more complex specification of the conditional mean, with the purpose of comparing volatility proxies through more elaborated models with non-linearities in the dynamics and 
(iii) the adoption of the proposed model for other applications of interest, e.g. for the analysis of spillover effects between market indices.

\section*{{Acknowledgements}}

The authors gratefully thank Reza Solgi for sharing the details of his previous research on univariate multiplicative error morels, Maria Conception Ausin for fruitful discussions on the sampling aspects of the proposed algorithm, Fabrizio Leisen, Antonio Canale, Bernardo Nipoti, Jim Griffin and Sonia Petrone for useful comments on intermediate results.

\bibliography{Bibliografia_UNITA}
\bibliographystyle{plainnat}

\end{document}